\documentclass[conf]{new-aiaa}

\usepackage{adjustbox,amsfonts,amsmath,array,bbm,bm,braket,cancel,centernot,enumerate,enumitem,fancyhdr,graphicx,hyperref,mathdots,mathtools,mdwlist,multirow,pgfplots,soul,subfigure,thmtools,tikz,tikz-qtree,titlesec}

\usepackage[utf8]{inputenc}

\hypersetup{colorlinks=true,linkcolor=blue,citecolor=blue,urlcolor=blue}    
\pgfplotsset{compat=1.18}
\allowdisplaybreaks 

\newcommand{\ketbra}[2]   {\left|#1\middle\rangle\!\middle\langle#2\right|}
\newcommand{\ot}{\otimes}
\newcommand\C{ {\mathcal C} }
\newcommand\D{ {\mathcal D} }
\newcommand\I{ {\mathcal I} }
\newcommand\N{ {\mathcal N} }
\newcommand\F{ {\mathcal F} }
\renewcommand\O{ \ensuremath{{\mathcal O}} }
\renewcommand\S{ {\mathcal S}}
\newcommand\W{ {\mathcal W} }
\newcommand{\eq}{\mathrm{eq}}
\newcommand{\ma}{\mathrm{Ma}}
\newcommand{\re}{\mathrm{Re}}
\newcommand{\rel}{\mathrm{rel}}

\renewcommand{\v}[1]{\ensuremath{\boldsymbol #1}}
\newenvironment{aligns}{\subequations \align} {\endalign \endsubequations}

\title{Simulating non-trivial incompressible flows \\with a quantum lattice Boltzmann algorithm}

\author{David Jennings\footnote{Quantum Applications Team, PsiQuantum, 700 Hansen Way, Palo Alto, CA 94304, USA}}
\author{Kamil Korzekwa*\footnote{Lead authors' emails: kkorzekwa@psiquantum.com and pmannix@psiquantum.com (authors are listed alphabetically within each affiliation).}}
\author{Matteo Lostaglio*}
\author{Paul Mannix*$^\dagger$}

\affil{PsiQuantum, 700 Hansen Way, Palo Alto, CA 94304, USA}
\author{Richard Ashworth\footnote{Research Scientist, Central Research and Technology, Airbus Operations Ltd, Pegasus House Aerospace Avenue, Filton, Bristol, UK}}
\author{Emanuele Marsili\footnote{Research Scientist, Central Research and Technology, Airbus Operations Ltd, Pegasus House Aerospace Avenue, Filton, Bristol, UK}}
\author{Stephen Rolston\footnote{Research Project Leader, Central Research and Technology, Airbus Operations Ltd, Pegasus House Aerospace Avenue, Filton, Bristol, UK}}
\affil{Airbus Operations Ltd, Pegasus House Aerospace Avenue, Filton, Bristol, UK}

\begin{document}

\maketitle

\begin{abstract}
    Quantum algorithms have been identified as a potential means to accelerate computational fluid dynamics (CFD) simulations, with the lattice Boltzmann method (LBM) being a promising candidate for realizing quantum speedups. Here, we extend the recent quantum algorithm for the incompressible LBM to account for realistic fluid dynamics setups by incorporating walls, inlets, outlets, and external forcing. We analyze the associated complexity cost and show that these modifications preserve the asymptotic scaling, and potential quantum advantage, of the original algorithm. Moreover, to support our theoretical analysis, we provide a classical numerical study illustrating the accuracy, complexity, and convergence of the algorithm for representative incompressible-flow cases, including the driven Taylor–Green vortex, the lid-driven cavity flow, and the flow past a cylinder. Our results provide a pathway to accurate quantum simulation of nonlinear fluid dynamics, and a framework for extending quantum LBM to more challenging flow configurations.
\end{abstract}


\section{Introduction}

In settings where the understanding and prediction of the behavior of fluid flow is the ultimate aim and, in particular, when experiments are either impossible (e.g., in oceanography and metrology) or prohibitively expensive/time-consuming (e.g., building and testing multiple prototypes in wind-tunnels), computational fluid dynamics (CFD) models have become indispensable~\cite{Spalart_Venkatakrishnan_2016}.
Due to the challenges associated with numerically modeling the non-trivial flow around a complex geometry with multiple different boundary conditions, a number of specialized high performance CFD codes have been developed. Albeit based on fundamentally different numerical methods including the finite-volume~\cite{patel2024assessing}, spectral-element~\cite{nek5000-web-page} and lattice Boltzmann method~\cite{bauer2021walberla}, these codes are highly optimized for parallelized execution on CPU and GPU architectures. As a result, it has been possible to simulate applications as diverse as landing-gear, wind farms and turbo-machinery, respectively. However, as these simulations seek to grow larger and better resolve the flow physics, computational constraints emerge. Notwithstanding the computational complexity increasing at least cubically with the Reynolds number $\re$, representing the flow field for a large computation demands also distributed memory. This, in turn, requires computational operations to be kept as local as possible in order to minimize the communication overhead~\cite{witherden2017future}. On this basis, tensor network approximations based on matrix product states~\cite{peddinti2024quantum}, super-resolution machine learning methods~\cite{page2025super} as well as quantum CFD algorithms~\cite{tennie2025quantum} have emerged.

Quantum computing offers a fundamentally different way to process information that is expected to provide transformative (in some cases exponential) improvements over what is possible with `classical' computing, i.e., computational devices (such as today's supercomputers) that process information via classical physics~\cite{feynman2018simulating, lloyd1996universal, berry2007efficient, aspuru2005simulated, mcardle2020quantum}. Instead of bits that are 0 or 1, a quantum computer uses quantum states that can explore many possibilities at once, and carefully designed algorithms `interfere' these possibilities to amplify the desired solution. Over the past decade, researchers have built algorithms that target tasks central to scientific computing, such as solving large linear systems~\cite{harrow2009quantum-5fb, costa2022optimal, jennings2023efficient}, estimating observables~\cite{danz2025calculating, kiani2022quantum,rendon2025exponentially}, and evolving dynamical models~\cite{berry2014high,montanaro2016quantum,berry2017quantum,jennings2024cost,an2023quantum, bagherimehrab2023fast,liu2023dense}. For nonlinear dynamics, ubiquitous in fluid flow, the nonlinearities must be put into a form amenable to a quantum computer. To achieve this, one can lift the nonlinear problem into a higher-dimensional \emph{linear} form (e.g., via embedding/expansion methods~\cite{carleman1932application}) and then solve the resulting system with quantum linear-algebra routines, which are known to circumvent the curse of dimensionality. Since the quantum-encoded data is a potentially exponentially large vector, the aim is not to print out the entire flow field, but to extract specific features or observables faster than classical methods might allow, especially for very large problem sizes. That said, potential speedups come with caveats: data must be encoded efficiently, the lifted system must be well-conditioned enough for the solver to work, and today’s hardware remains noisy, so most proposals target future, fault-tolerant machines~\cite{liu2021efficient, an2022efficient,krovi2023improved, costa2025further, wu2025quantum, jennings2025quantum}. Nevertheless, emerging quantum algorithms are a promising tool-in-the-making for selected, structured nonlinear problems, which are expected to supplement existing CFD methods in novel and impactful ways.

One prominent example of a structured nonlinear problem is given by the \emph{lattice Boltzmann equation} (LBE), whose nonlinear strength scales with the Mach number $\ma$, and whose memory demands are more intensive as compared with those of other CFD methods. On this basis, a number of quantum algorithms employing Carleman embedding/expansion methods have been proposed in recent years~\cite{li2025potential,penuel2024feasibility,sanavio2024lattice,turro2025practical}. In Ref.~\cite{jennings2025end-to-end}, it was shown that these prior works did not properly address the crucial convergence of the Carleman embedding, which is essential for resolving nonlinear physics properly. The work also highlighted that prior algorithms suffered from inefficient information-extraction from the quantum state, as well as other algorithmic bottlenecks, such as unreasonably small time-step requirements. In order to circumvent these obstacles, the authors of Ref.~\cite{jennings2025end-to-end} developed a quantum algorithm for the \emph{incompressible} lattice Boltzmann equation~\cite{he1997lattice, guo2000lattice} and provided evidence that a modest quantum advantage can occur over classical methods, albeit with caveats that require further study and development. 

While Ref.~\cite{jennings2025end-to-end} concentrated on building the theoretical framework for the quantum lattice Boltzmann algorithm, connecting these results to real-world problems of industrial relevance requires incorporating non-trivial boundary conditions and external driving. In this paper, we present the modifications required to extend Ref.~\cite{jennings2025end-to-end} so that walls, inlets, outlets, and forcing can be accommodated. We explain the corresponding changes to the quantum algorithm, assess their complexity cost, and argue that the introduced extensions should not significantly slow down the designed quantum algorithm. In addition, we conduct a thorough numerical assessment of the Carleman convergence to the correct nonlinear dynamics. This is performed by computing solutions of the driven Taylor-Green vortex, the lid-driven cavity flow, and the flow past a cylinder.

The structure of this work is as follows. First, in Sec.~\ref{sec:framework}, we briefly review the incompressible LBM and its associated quantum algorithm developed in Ref.~\cite{jennings2025end-to-end}, introducing the necessary notation. Then, in Sec.~\ref{sec:extensions}, we explain how, and at what complexity cost, one can extend this algorithm to include non-trivial boundary conditions in the form of walls, inlets and outlets, as well as how to add driving. In Sec.~\ref{sec:case_studies}, we present the convergence and complexity analysis for exemplary CFD settings. Finally, Sec.~\ref{sec:conclusions} contains conclusions and outlook for future work, while the Appendix contains extra technical details to make this work self-contained. 


\section{Existing framework}
\label{sec:framework}


\subsection{Lattice Boltzmann method}

The \emph{lattice Boltzmann equation} (LBE) is a kinetic model of fluid dynamics that
describes the fluid mesoscopically~\cite{kruger2016lattice}. It conserves mass and momentum by construction and its low-order moments recover the Navier-Stokes equations in the regime where the characteristic length scales of the problem are much larger than the molecular mean path. Within LBE, the fluid is modeled at time $t$ and position $\v{r}$ in $D$-dimensional space by a vector-valued function $\v{g}(\v{r},t)$, whose components $g_m(\v{r},t)$, with $m\in\{1,\dots,Q\}$, are proportional to the probability density of finding a fluid particle with a discrete velocity $\v{e}_m$. LBE is derived from the \emph{discrete Boltzmann equation} (DBE) describing continuous dynamics via:
\begin{align}
    \label{eq:DBE}
    \partial_t g_m(\v{r},t) +\v{e}_m\cdot \nabla g_m(\v{r},t) = \Omega_m(\v{r},t),
\end{align}
where $\v{e}_m\cdot \nabla g_m$ is an advection term that describes \emph{streaming}, i.e., the flow of the fluid in space, while the collision operator $\Omega_m$ describes \emph{collisions} that are local in space. We adopt the Bhatnagar-Gross-Krook (BGK) operator~\cite{bhatnagar1954model}:
\begin{equation}
    \label{eq:BGK}
    \Omega_m(\v{r},t) = -\frac{1}{ \tau} \left[g_m(\v{r},t)-g_m^\eq(\v{r},t)\right],
\end{equation}
where $\tau$ is the relaxation time determining the speed of equilibration, and $\v{g}^\eq$ is the Taylor expansion of the local Maxwell equilibrium up to second order terms, which for an incompressible flow is given by~\cite{he1997lattice}: 
\begin{align}
    \label{eq:eq}
    g_m^\eq(\v{r},t)= \frac{p}{c_s^2} \; w_m + \rho_0 \left( \frac{\v{e}_m\cdot \v{u}(\v{r},t)}{c_s^2} + \frac{(\v{e}_m\cdot \v{u}(\v{r},t))^2}{2c_s^4} -\frac{|\v{u}(\v{r},t)|^2}{2c_s^2}\right) w_m + O(\ma^3).
\end{align}
In the above, $\rho_0$ denotes the constant fluid density, $c_s$ the speed of sound, $\ma:=|\v{u}|/c_s$ the Mach number, and $\v{w}$ is a vector of Maxwell weightings for the different discrete velocities satisfying $w_m\geq 0$ and \mbox{$\sum_m w_m=1$}. The pressure $p$ and velocity vector $\v{u}$ of the fluid are recovered as moments via:
\begin{equation}
    \label{eq:p_and_u}
    p(\v{r},t):= c_s^2 \sum_{m=1}^Q g_m(\v{r},t),\qquad
    \v{u}(\v{r},t):= \frac{1}{\rho_0}\sum_{m=1}^Q g_m(\v{r},t)\v{e}_m.
\end{equation}
As shown in Ref.~\cite{he1997lattice}, this formulation recovers the Navier-Stokes equations, albeit with the divergence-free condition approximated as $\nabla \cdot \v{u} = \frac{1}{c_s^2} \partial_t p \sim O(\ma^2)$.\footnote{Although this is suitable for steady or slowly varying flows, the collision operator proposed in Ref.~\cite{guo2000lattice} (also quadratic in $g_m$) is by contrast exact and therefore preferable for time-dependent flows.} Therefore, at the expense of discarding weak compressibility effects, which for $\ma \ll 1$ are already negligible, an $O(\ma^2)$ incompressible LBE formulation whose collision integral is quadratic in $g_m$ is obtained. 

To obtain the numerical solution of Eqs.~\eqref{eq:DBE} - \eqref{eq:p_and_u}, the system must be discretized in space, $r_i \in [\Delta x,2\Delta x, \dots, N_i \Delta x]$, and in time, $t \in [0,\Delta t,2\Delta t, \dots, N_t \Delta t]$. Integrating the left hand side of Eq.~\eqref{eq:DBE} using the method of characteristics and approximating the integral on the right side of Eq.~\eqref{eq:DBE} using the trapezoidal rule, one obtains:
\begin{equation}
    \label{eq:LBE_incomp}
    g_m(\v{r}+\v{e}_m\Delta t,t+\Delta t) = g_m(\v{r},t) - \frac{\Delta t}{\tau + \Delta t/2}[g_m(\v{r},t)-g^{\mathrm{eq}}_m(\v{r},t)],
\end{equation}
which is second order accurate in both space and time~\cite{kruger2016lattice}. To determine how Eq.~\eqref{eq:LBE_incomp} relates to the macroscopic Navier-Stokes equations, Chapman-Enskog analysis must be applied, which shows that the kinematic shear viscosity,
\begin{equation}
    \nu = {c_s}^2\left(\tau -\frac{\Delta t}{2}\right),
    \label{eq:nu}
\end{equation}
is a function of the relaxation time, time-step and speed of sound. As discussed in Ref.~\cite{kruger2016lattice}, the factor of $(\tau - \Delta t/2)$ appearing in Eqs.~\eqref{eq:LBE_incomp} and~\eqref{eq:nu} is not general, but rather a consequence of the time-discretization employed. Because we have used a second-order time discretization, the relaxation time $\tau$ is shifted by $\Delta t/2$. Although physically significant length- and time-scales are typically chosen, the system from Eqs.~\eqref{eq:eq} - \eqref{eq:nu} is instead made non-dimensional by choosing the lattice spacing $\Delta x$ and time step $\Delta t$ (also known as lattice units) along with the reference density $\rho_0$, such that:
\begin{align}
    t=t^\star \Delta t,\quad T = N_t \Delta t,\quad \tau= \tau^\star \Delta t,\quad
    \v{r}=\v{r}^\star\Delta x,\quad L_i = N_i \Delta x,\quad
   \v{e}_m=\v{e}_m^\star \frac{\Delta x}{\Delta t},\quad \v{u}=\v{u}^\star \frac{\Delta x}{\Delta t}, \quad p = p^\star c_s^2 \rho_0,
\end{align}
where superscript star denotes a non-dimensional variable, $L_i$ is the length of the $i^{th}$ side of the domain and $T$ is the size of time-interval considered. This choice restricts $t^\star\in\{0,\dots N_t\}$ and $r_i^\star\in\{1,\dots, N_i\}$ to integer values, such that for an incompressible isothermal flow $c_s = \frac{1}{\sqrt{3}}\frac{\Delta x}{\Delta t}$ and $\v{e}_m^\star, \, w_m$ are given by Table~\ref{table:D2Q9} in the Appendix.  

Anticipating the quantum formulation of Eq.~\eqref{eq:LBE_incomp}, the system state at each time step $t^\star$ can be represented as:
\begin{equation}
    \v{g}(t^\star) = \sum_{\v{r}^\star} \sum_{m=1}^Q  g_m(\v{r}^\star,t^\star) \ket{\v{r}^\star} \ot \ket{m},
\end{equation}
where $\ot$ denotes the tensor product and $\ket{\v{r}^\star}, \ket{m}$ denote position and velocity registers, respectively, using the Dirac ket notation. The dimension of $\v{g}$ is $d=NQ$, where $N$ is the total number of spatial lattice points. Equation \eqref{eq:LBE_incomp} can be then be divided into collision (C) and streaming (S) steps, and rewritten in a matrix form as: 
\begin{aligns}
    \label{eq:sLBE_col_mat}
    \v{g}^C(t^\star) & = \left(  I+ F_1\right)\v{g}(t^\star)+ F_2 \v{g}(t^\star)^{\ot 2},\\
    \label{eq:sLBE_str_mat}
    \v{g}(t^\star+1) & = S\v{g}^C(t^\star),
\end{aligns}
where $F_k$ are $d\times d^k$ matrices describing linear and quadratic contributions from collisions and $S$ is a $d\times d$ permutation matrix encoding streaming, see the Appendix for details.


\subsection{Carleman embedding of the nonlinear dynamics}

The lattice Boltzmann equation is nonlinear, whereas quantum algorithms running on quantum computers are built from linear (unitary) operations, so there is a structural mismatch. The way in which this mismatch is resolved is by embedding the nonlinear dynamics into a much higher-dimensional linear system, whose variables are functions of the original ones. While this lifted system can be overwhelmingly large for classical solvers, quantum circuits act on superpositions and apply linear operators implicitly, so the cost need not scale with the full dimension in the same way as explicit classical representations. In practice, this makes it viable to work with high-dimensional embeddings on a quantum computer, provided the embedding can be implemented efficiently and remains well-behaved.

In order for the incompressible LBE problem, captured by Eqs.~\eqref{eq:sLBE_col_mat}-\eqref{eq:sLBE_str_mat}, to be amenable for the quantum computer, we use the Carleman embedding technique~\cite{carleman1932application}. Instead of representing the state of the fluid at each moment in time by a $d$-dimensional vector $\v{g}$, it is now represented by the \emph{Carleman vector}:
\begin{equation}
    \v{y}(t^\star)  := [\v{y}_1(t^\star),\v{y}_2(t^\star),\dots,\v{y}_{N_C}(t^\star)], \qquad  \v{y}_k(0) = \v{g}^{\otimes k} (0)
    \label{eq:truncatedCarlemanstate}
\end{equation}
of dimension
\begin{equation}
    \label{eq:Carleman_dimension}
    d_C:= d+d^2+\dots +d^{N_C} = \frac{d(d^{N_C}-1)}{d-1},
\end{equation}
which carries information about all powers of the incompressible LBE evolution up to the $N_C$-th power. Here, $N_C$ is the finite Carleman truncation order leading to Carleman truncation error $\epsilon_C$, i.e., the evolution obtained from a truncated linear system will generally differ from the original LBE dynamics.

The update rules for a $d$-dimensional vector $\v{g}(t^\star)$ from Eqs.~\eqref{eq:sLBE_col_mat}-\eqref{eq:sLBE_str_mat},
can now be approximated by a linear update rule for a $d_C$-dimensional vector $\v{y}(t^\star)$~\cite{jennings2025end-to-end}:
\begin{equation}
    \label{eq:LBE_recurrence}
    \v{y}(t^\star+1) = \S\C \v{y}(t^\star),
\end{equation}
where $\S$ and $\C$ are $d_C\times d_C$ Carleman streaming and collision matrices, see the Appendix for details. The above recurrence relation has a clear solution given by:
\begin{equation}
    \label{eq:carlemanSolution}
    \v{y}(t^\star) = (\S\C)^{t^\star} \v{y}_{\mathrm{ini}}, \qquad \v{y}_{\mathrm{ini}} = [\v{g}(0),\v{g}(0)^{\ot 2},\dots,\v{g}(0)^{\ot N_C}].
\end{equation}
Therefore, for the initial state $\v{g}(0)$, the state of the system after $N_t$ streaming and collision steps is approximated by:
\begin{equation}
    \v{g}(N_t) \approx_{\epsilon_C} \left[(\S\C)^{N_t} \v{y}_{\mathrm{ini}}\right]_1,
\end{equation}
where subscript 1 indicates the first block of the Carleman vector.

Next, the LBE problem with $N_t$ streaming and collision steps can be written in the form of the following linear system of dimension $d_C(N_t+1)$:
\begin{align}
    \label{eq:AH}
        \begin{pmatrix}
            I& 0 & 0 & \dots&0&0\\
            -\S\C   & I& 0 & \dots&0&0\\
            0     & -\S\C    & I&\dots&0&0\\
            \vdots&\vdots &  \vdots& \ddots&\vdots&\vdots \\
            0&0&0&\dots&I&0\\
            0&0&0&\dots&-\S\C&I
        \end{pmatrix}
        \begin{pmatrix}
            \v{y}(0)\\
            \v{y}(1)\\
            \v{y}(2)\\
            \vdots\\
            \v{y}(N_t-1)\\
            \v{y}(N_t)
        \end{pmatrix} = 
        \begin{pmatrix}
            \v{y}_{\mathrm{ini}}\\
            0\\
            0\\
            \vdots\\
            0\\
            0
        \end{pmatrix},
\end{align}
or in a more compact form as
\begin{equation}
    \label{eq:linearHistory}
    A\v{Y}=\v{b}.    
\end{equation}
The vector $\v{Y}$ is the \emph{Carleman history state}, carrying information about Carleman vector at all times during the simulated evolution:
\begin{equation}
    \v{Y} = A^{-1} \v{b} = \left(\v{y}_{\mathrm{ini}},\S\C \v{y}_{\mathrm{ini}}, (\S\C)^2 \v{y}_{\mathrm{ini}},\dots, (\S\C)^{N_t-1} \v{y}_{\mathrm{ini}}, (\S\C)^{N_t} \v{y}_{\mathrm{ini}} \right)^\top.
    \label{eq:YH}
\end{equation}
The problem therefore reduces to performing a matrix inversion, and outputting the solution to the above set of linear equations. Although the solution vector $\v{Y}$ can be exponentially large, as a function of the Carleman truncation scale, one can now exploit the particular features of a quantum computer that allow one to process such vectors efficiently. More precisely, it is known that \emph{quantum linear solvers} (QLS) have exponential speedups over classical methods for certain structured problems~\cite{harrow2009quantum-5fb} and, more recently, it has been shown that they can also provide exponential speedups in the context of simulating nonlinear dynamical systems via the above Carleman method~\cite{jennings2025quantum}.


\subsection{Solving exponentially large linear systems on a quantum computer}

Quantum algorithms for solving systems of linear equations as in Eq.~\eqref{eq:linearHistory} typically return the solution in an encoded (quantum) form rather than a full list of numbers~\cite{harrow2009quantum-5fb, costa2022optimal, jennings2023efficient, lin2022lecture}, so they are most useful when one only needs certain features, like averages or correlations, rather than every entry. Their performance also depends on how well-behaved the equations are (in terms of the condition number $\kappa_A = \|A\|\cdot \|A^{-1}\|$) and on whether the problem has structure that allows one to load data and apply matrices efficiently. It is rigorously established that promising quantum speedups exist, but only for the right kinds of large, structured problems~\cite{harrow2009quantum-5fb}.

The ability of quantum linear solvers to provide large speedups relies crucially on the output of a quantum computer, which differs radically from what one has from a classical computer. This is because the solution $\v{Y} = A^{-1}\v{b}$ is encoded in a \emph{quantum state}. Physically, this corresponds to a quantum mechanical system (such as a system of photons, superconductors, atoms, etc.) being in a superposition of two or more classical states, e.g., a particle being in a superposition of two locations $A$ and $B$. Mathematically, we write this superposition state in the Dirac notation as a vector $\ket{\psi} = a \ket{A} + b\ket{B}$, where $a$ and $b$ are complex numbers that obey $|a|^2+|b|^2=1$. Within quantum theory, the classical states are represented as orthonormal basis vectors, and so in this basis we have that $\ket{\psi}$ is the vector with components $(a,b)$. More generally, we label classical states with integers, and a general $d$-dimensional quantum state takes the form:
\begin{equation}
    \ket{\psi} = a_0 \ket{0} + a_1 \ket{1} + \cdots + a_{d-1} \ket{d-1},
\end{equation}
with the condition $\sum_j |a_j|^2=1$. If we measure the system, we collapse the quantum state probabilistically onto one of the states $\ket{k}$, and the probability that this occurs is $p_k = |a_k|^2$. Therefore, the normalization condition is simply the condition that measurement probabilities in quantum theory sum to $1$. Upon getting outcome $k$, the quantum state collapses $\ket{\psi} \rightarrow \ket{k}$. At the level of linear algebra, this can be described (up to normalizing the vector) by the application of a projection matrix:
\begin{equation}
    (a_0,\dots, a_{d-1}) \rightarrow P_k  (a_0,\dots, a_{d-1}) = (0,0, \dots,0, a_k,0,\dots ,0).
\end{equation}
Here, $P_k$ is the projection matrix onto the $k$-th basis vector that can be written in the Dirac notation as $P_k = \ketbra{k}{k}$. 

More generally, we have the compact Dirac notation convention defining rank-1 matrices $\ketbra{m}{k}$ given by:
\begin{equation}
    \ketbra{m}{k} \sum_{j} a_j \ket{j} = \sum_{j} a_j \ketbra{m}{k} j\rangle = a_k \ket{m},
\end{equation}
in other words $\bra{k}j\rangle = \delta_{j,k}$ for any $k,j$. Given two quantum states $\ket{\psi} = \sum_j a_j \ket{j}$ and $\ket{\phi} = \sum_j b_j \ket{j}$, we then define the conjugate transposed vector $\bra{\phi} := \sum_j b^*_j \bra{j}$, the matrix 
\begin{equation}
    \ketbra{\psi}{\phi} = \sum_{j,k} a_j b^*_k\ketbra{j}{k},
\end{equation}
and the complex number obtained from taking the inner product of these two vectors:
\begin{equation}
    \bra{\phi}\psi \rangle = \sum_{j,k}a_j b^*_k \delta_{j,k} = \sum _{k} a_k b^*_k.
\end{equation}
We also can have multiple copies of quantum states and denote, e.g., two systems with the first in the state $\ket{\psi}$ and the second in the state $\ket{\phi}$ as $\ket{\psi}\otimes \ket{\phi}$. In other words: multiple copies are denoted with terms separated by tensor products, and $A^{\otimes n} = A\otimes A \otimes \cdots \otimes A$ denotes $n$ copies of $A$. In this setting, the rules of quantum theory correspond to linear transformations (matrices on vectors) and linear algebra manipulations (compositions of matrices, inner products, tensor products, etc.).
 
To implement quantum algorithms on a quantum computer, every component must be encoded in a \emph{quantum circuit} (a sequence of instructions as to how to manipulate the quantum system). Mathematically, a quantum circuit is described by an exponentially large unitary matrix that acts on quantum states. This exponential matrix could describe various data, such as vectors and matrices. For example, the primitive of \emph{state preparation} involves a quantum circuit $U_{\v{b}}$ that acts on a fiducial vector $\ket{0} = (1,0,0\dots, 0)$ and prepares the new vector $\ket{\v{b}} = U_{\v{b}} \ket{0} =\frac{1}{\|\v{b}\|} (b_0,b_1, \dots b_{d-1})$. In other words, the first column of $U_{\v{b}}$ is simply the normalized state vector $\ket{\v{b}}$. Another key ingredient is the encoding of an arbitrary matrix $A$ inside a unitary circuit. In order to respect unitarity, the matrix $A$ is typically rescaled $A \rightarrow A/\alpha_A$ by some positive constant $\alpha_A \geq\|A\|$, and then a quantum circuit $U_A$ is constructed such that the top-left block of the matrix is identical to $A/\alpha_A$; this is called a \emph{unitary block-encoding} of the matrix $A$.

A standard approach to solving a linear system $A \v{Y} = \v{b}$ on a quantum computer~\cite{harrow2009quantum-5fb, costa2022optimal, lin2022lecture} assumes access to (i) an efficient \emph{unitary state preparation} $U_{\v{b}}$ such that $U_{\v{b}}\ket{0}=\ket{\v{b}}$, and (ii) a \emph{block-encoding} $U_A$ of $A$, i.e., a unitary $U_A$ acting on $\log d +n_A$ qubits whose top-left block equals $A/\alpha_A$
for some \emph{block-encoding prefactor} $\alpha_A \ge  \|A\|$.
Given $(U_{\v{b}},U_A)$, modern \emph{quantum linear solvers} output  a state proportional to $A^{-1} \ket{\v{b}}$ as a normalized quantum state. Optimal quantum linear solvers prepare the solution $\ket{\v{Y}}$ within error $\epsilon$ using $O\!\left(\alpha_A\,\kappa_A\,\log\!\frac{1}{\epsilon}\right)$ applications of $U_A,U^\dagger_A$ and of the preparation unitaries $U_{\v{b}}, U_{\v{b}}^\dagger$, where $\kappa_A$ is the  condition number of $A$ on the relevant subspace.
The dependence on both the condition number and the block-encoding prefactor is linear; thus large $\alpha_A$ directly inflates runtime. In practice, (quantum or classical) preconditioning can reduce $\kappa_A$, but any gains must be balanced against changes in $\alpha_A$ and additional oracle costs.

One final subtlety exists for QLS: the output. The output of the quantum linear solver applied to the Carleman system will be a \emph{quantum state} $\ket{\v{Y}}$ of the quantum computer. This is quite distinct from the output of a classical computer, which can, e.g., print the data on a screen to be perused. Instead, the quantum output $\ket{\v{Y}}$ is an exponentially large vector that is `stored' in the quantum computer (e.g., for a quantum computer with $1000$ qubits, the vector has dimension $2^{1000} \approx 10^{300}$). In order to extract meaningful information, one must perform a carefully chosen measurement that extracts a much smaller amount of information from $\ket{\v{Y}}$. The full end-to-end quantum simulation must therefore build in the exact information that is required into the full pipeline, and the information extraction becomes itself a potential source of inefficiency. While we do not discuss that in this work, we refer the interested reader to Ref.~\cite{jennings2025end-to-end}, where it is shown how, and at what complexity cost, one can extract the drag force on an obstacle from a quantum state $\ket{\v{Y}}$.


\section{Extensions to nontrivial flows and geometries}
\label{sec:extensions}


Having introduced the incompressible Carleman LBM algorithm from Ref.~\cite{jennings2025end-to-end} and discussed how to recover its solution using QLS, we now extend it to accommodate non-trivial geometries and driving via a body forcing or moving boundaries. These extensions will allow us to investigate the performance of our quantum algorithm on benchmark CFD problems, such as the forced Taylor-Green vortex, the lid-driven cavity flow and the flow past a cylinder, considered in Sec.~\ref{sec:case_studies}.

\subsection{Geometry}
\label{sec:Geometry}


\subsubsection{Extending the model}

For periodic boundary conditions, there are only fluid nodes and the system can be encoded in the streaming matrix:
\begin{equation}
    \label{eq:streaming_per}
    S = \sum_{\v{r}^\star}\sum_{m=1}^Q \ketbra{\v{r}^\star+\v{e}_m^\star}{\v{r}^\star} \ot \ketbra{m}{m},
\end{equation}
which translates populations $g_m$ from position $\v{r}^\star$ and with a discrete velocity $m$ to the next spatial position $\v{r}^\star+\v{e}_m^\star$, where the addition should be interpreted as modulo the number of lattice points per direction, thus encoding the periodic domain. We want to modify this streaming matrix in order to account for the no-slip (Dirichlet) and free-slip (Neumann) boundary conditions required to consider domains with walls, inlets and outlets. 

The effect of a Dirichlet boundary condition can be modeled via the bounce-back rule following Ref.~\cite{kruger2016lattice}, whereby one prescribes these macroscopic boundary conditions via the populations as:
\begin{equation}
    g_{\bar{m}}(\v{r}^\star_b,t^\star) = g^C_m(\v{r}^\star_b,t) - 2 w_m \frac{\v{e}^\star_m \cdot \v{u}^\star_w}{ {c^{*}_s}^2 }.
    \label{eq:bounce_back}
\end{equation}
Here, $\v{r}^\star_b$ is the boundary node, $g^C_m$ the post-collision population, $\bar{m}$ the reflection of/population opposite $m$, and $\v{u}^\star_w$ the velocity vector of the wall located at $\v{r}^\star_w = \v{r}^\star_b + \frac{1}{2}\v{e}^\star_m$. As the second term in Eq.~\eqref{eq:bounce_back} accounts for the motion imparted by a moving wall, we will include it in the driving matrix described in section \ref{sec:driving}. The effect of a Neumann boundary condition can be modeled via the non-equilibrium extrapolation scheme following Refs.~\cite{junk2008outflow, guo2013lattice}. This scheme provides the populations in terms of their equilibrium and non-equilibrium components at the boundary via:
\begin{equation}
    g_m(\v{r}^\star_b, t^\star) = g^{\eq}_m(p^\star_b, \v{u}^\star_b) + \left( g_m(\v{r}^\star_f, t) - g^{\eq}_m(p^\star_f, \v{u}^\star_f) \right),
\end{equation}
where $\v{r}^\star_f$ is one node to the interior of the boundary and $(p^\star_f, \v{u}^\star_f)$ its corresponding pressure and velocity. For a Neumann-boundary condition at which we want to enforce $\partial_x \v{u}^\star = \partial_x p^\star = 0$, for example, we can set $(p^\star_b, \v{u}^\star_b) = (p^\star_f, \v{u}^\star_f)$, such that the scheme reduces to prescribing populations for the unknown $g_m$ at $\v{r}^\star_b$ by copying them from $\v{r}^\star_f$.

To illustrate how these boundary conditions can be incorporated into the streaming matrix, we consider a domain bounded by a wall region, an inlet region and outlet region, all with particular geometries. 
Every node is either an inlet node, an outlet node, a wall node, or otherwise a fluid node interior to the simulation region. We introduce the following indicator functions on the discrete phase space $\{(\v{r}^\star, m)\}$ that encode the boundary regions within the model:
\begin{aligns}
    \label{eq:walls}
    \W_{\v{r}^\star} &= \left\{
    \begin{array}{ll}
         1&:~\v{r}^\star \mathrm{~is~a~wall~node,}   \\
         0& :\mathrm{~otherwise,}
    \end{array}
    \right.\\
    \label{eq:inlet}
    \I_{\v{r}^\star,m} &= \left\{
    \begin{array}{ll}
        1&:~\v{r}^\star\mathrm{~is~an~inlet~node~and~}\v{r}^\star+\v{e}^\star_m \mathrm{~is~outside~the~simulation~region,}   \\
         0& :~\mathrm{otherwise,}
    \end{array}
    \right.\\
    \label{eq:outlet}
    \O_{\v{r}^\star,m} & = \left\{
    \begin{array}{ll}
        1&:~\v{r}^\star\mathrm{~is~an~outlet~node~and~}\v{r}^\star+\v{e}^\star_m \mathrm{~is~outside~the~simulation~region,}   \\
         0& :~\mathrm{otherwise.}
    \end{array}
    \right.
\end{aligns}

At the walls we impose a no-slip Dirichlet condition such that the effect of the first term in Eq.~\eqref{eq:bounce_back} can be achieved via the following modification of matrix elements of the streaming matrix:
\begin{align}
     \label{eq:streaming_walls}
    S_{\mathrm{wall}} = &\sum_{\v{r}^\star}\sum_{m=1}^Q (1-\W_{\v{r}^\star+\v{e}_m^\star} )(1-\W_{\v{r}^\star})\ketbra{\v{r}^\star+\v{e}_m^\star}{\v{r}^\star} \ot \ketbra{m}{m}\\
    & + \sum_{\v{r}^\star}\sum_{m=1}^Q \W_{\v{r}^\star+\v{e}_m^\star} (1-\W_{\v{r}^\star})\ketbra{\v{r}^\star}{\v{r}^\star} \ot \ketbra{\bar{m}}{m}\\
    & + \sum_{\v{r}^\star}\sum_{m=1}^Q \W_{\v{r}^\star} \ketbra{\v{r}^\star}{\v{r}^\star} \ot \ketbra{m}{m}.
\end{align}
In the above, the first term corresponds to unmodified streaming when both nodes considered are fluid; the second term encodes bouncing, when a particle in a fluid node hits a wall node; and the third one simply means no evolution (identity matrix) for the wall nodes. We note that the streaming matrix modified according to the above recipe is still a permutation matrix, so $S_{\mathrm{wall}}$ stays unitary and therefore can be realized directly by a quantum circuit. 

At an outlet, we impose a Neumann-boundary condition. To account for this condition in the streaming matrix, we introduce one more correction of the following form:
\begin{equation}
    S_{\mathrm{Neumann}} = I + \sum_{\v{r}^\star} \sum_{m=1}^Q \O_{\v{r}^\star,m} \ketbra{\v{r}^\star}{\N(\v{r}^\star)} \ot \ketbra{\bar{m}}{\bar{m}},
\end{equation}
and consider $S_{\mathrm{Neumann}} S_{\mathrm{wall}}$ instead of $S_{\mathrm{wall}}$. This prescribes to simply replace the missing outlet occupations after streaming at position $\v{r}^\star$ with a value copied from the neighboring interior node at position $\N(\v{r}^\star)$, one lattice node away along the normal to the outlet towards the simulation region\footnote{For example, in a two-dimensional setting with a vertical outlet being given by the points $(N_x,y)$ with $y\in\{1,\dots,N_y\}$ and $N_x$ the horizontal location of the outlet, we simply have $\N(\v{r}^\star)=\v{r}^\star-(1,0)$.}.

To model the edges of the simulation regions at an inlet and outlet, we also introduce the following correction term:
\begin{equation}
    S_{\mathrm{edge}} = \sum_{\v{r}^\star} \sum_{m=1}^Q (1-\I_{\v{r}^\star,m})(1-\O_{\v{r}^\star,m}) \ketbra{\v{r}^\star}{\v{r}^\star} \ot \ketbra{m}{m},
\end{equation}
and consider $S_{\mathrm{wall}}S_{\mathrm{edge}} $ instead of $S_{\mathrm{wall}}$. This term ensures that the fluid can go outside the simulation region via an inlet and an outlet, instead of moving periodically to the other side of the region. The final streaming term that accounts for walls, inlets and outlets is thus given by:
\begin{equation}
    S' = S_{\mathrm{Neumann}}S_{\mathrm{wall}}S_{\mathrm{edge}}.
\end{equation}
This linear map is fully defined once we specify the indicator functions for the wall nodes, the inlet nodes, and the outlet nodes. If one can efficiently encode these three regions in a sufficiently compact functional description, then one can construct a block-encoding of $S'$, which governs the linearized dynamics of the fluid. For simplicity, we shall assume straight boundaries aligned with the lattice nodes, with inlet and outlet nodes forming orthogonal lines to the walls at either end of the desired simulation region. 


\subsubsection{Algorithmic extensions}

As described in detail in Ref.~\cite{jennings2025end-to-end}, the workhorse of our quantum algorithm for LBE is the quantum linear solver of Ref.~\cite{dalzell2024shortcut} applied to the linear system from Eq.~\eqref{eq:AH}. This solver takes as its inputs the state preparation unitary $U_{\v{b}}$ and the unitary block-encoding $U_A$ of the linear system $A$. Since the modifications to the streaming matrix described in the previous section changes $A$, we will now explain how to block-encode the required linear system in a way that incorporates the wall boundary and inlet/outlet physics.

Let us denote by $A'$ the linear system using a modified streaming matrix $S'$, and by $A$ the original system with periodic $S$. Let us also recall that with an appropriately chosen tensor encoding~\cite{jennings2025end-to-end}, the Carleman streaming matrices can be represented by $\S=S^{\ot N_C}$ and $\S'=S'^{\ot N_C}$. In Ref.~\cite{jennings2025end-to-end}, it was shown that the block-encoding prefactor $\alpha_A$ and the number of ancillary qubits needed $n_A$ are given by:
\begin{equation}
    \alpha_{A} = \alpha_\C + 1,\qquad n_{A} = n_\C+2,
\end{equation}
where $\alpha_\C$ and $n_\C$ are the block-encoding parameters for the Carleman collision matrix. This result was obtained under the assumption that $\S$ is unitary (so that $\alpha_{\S}=1$ and $n_\S=0$). However, it is straightforward to show that for a general matrix, such as $\S'$, the expressions for the block-encoding parameters become:
\begin{equation}
    \alpha_{A'} = \alpha_{\S'} \alpha_\C + 1,\qquad n_{A} = n_{\S'} + n_\C+2.
\end{equation}
We can relate these parameters to the block-encoding parameters for the modified streaming term for a single-copy of the system. In particular, if we can construct a block-encoding of $S'$ with block-encoding prefactor $\alpha_{S'}$ using $n_{S'}$ ancillary qubits, the linear system $A'$ could be block-encoded with:
\begin{equation}
    \alpha_{A'} = (\alpha_{S'})^{N_C} \alpha_\C + 1,\qquad n_{A'} = N_C n_{S'} + n_\C+2.
\end{equation}

In order to achieve this, we will employ the following construction. First, recall that a partial permutation matrix is a matrix in which each row and each column contains at most one 1, and all other entries are 0. Given a partial permutation matrix $\tilde{\Pi}$, one can always complete it to a permutation matrix $\Pi$. These two matrices then satisfy:
\begin{equation}
    \label{eq:pi_property}
    \tilde{\Pi}\tilde{\Pi}^\dagger = \tilde{\Pi}\Pi^\dagger =  \Pi\tilde{\Pi}^\dagger = \tilde{I}, 
\end{equation}
where $\tilde{I}$ is the identity matrix on the image of $\tilde{\Pi}$. One can then block-encode $\tilde{\Pi}$ with a unit block-encoding prefactor, $\alpha_{\tilde{\Pi}}=1$, using a single ancillary qubit via:
\begin{equation}
    U_{\tilde{\Pi}} = \tilde{\Pi} \ot I + (\tilde{\Pi} - \Pi) \ot X,
\end{equation}
where we used a Pauli $X$ unitary on the ancillary qubit. Indeed, using Eq.~\eqref{eq:pi_property}, one can verify that $U_{\tilde{\Pi}}$ is unitary and that
\begin{equation}
    (I\ot \bra{0}) U_{\tilde{\Pi}} (I \ot \ket{0}) = \tilde{\Pi}.
\end{equation}

Now, we observe that $S_{\mathrm{edge}}$ is a partial permutation matrix, whereas $S_{\mathrm{Neumann}}$ is a sum of two partial permutation matrices. Hence, using the block-encoding described above together with a linear combination of unitaries (LCU)~\cite{lin2022lecture} applied to two components of $S_{\mathrm{Neumann}}$, we can straightforwardly construct the block encoding of $S'$ with a block-encoding prefactor $\alpha_{S'}=2$ using $n_{S'}=3$ ancillary qubits. We thus conclude that:
\begin{equation}
    \label{eq:A_prime_BE}
    \alpha_{A'} = 2^{N_C} \alpha_\C + 1 \approx 2^{N_C} \alpha_A,\qquad n_{A'} = n_A + 3N_C,
\end{equation}
so that incorporating non-trivial geometries into our quantum algorithm for LBE increases the query complexity by a factor of $2^{N_C}$ and requires additional $3N_C$ ancillary qubits. This additional cost should not have a significant effect on the efficiency of the algorithm, as it is expected that relatively low $N_C$ should suffice for simulations. More precisely, using the Carleman truncation error model~\cite{jennings2025end-to-end}, which predicts exponential damping of the Carleman error as $\epsilon_C\propto 2^{-|\Gamma| N_C}$ (as long as one is within the convergence radius), the increase in the query complexity is scaling as $O(1/\epsilon_C^{1/|\Gamma|})$. For example, for the lid-driven cavity flow discussed in Sec.~\ref{sec:lid}, one can estimate $|\Gamma|$ to be approximately $4.19$ for $\re=20$, which translates into an increase in the query complexity scaling as $O(1/\epsilon_C^{0.24})$. 

Also, note that the complexity increase discussed above is due to a modified block-encoding prefactor, $\alpha_{A'}$ instead of $\alpha_A$, but the query complexity can also increase due to the dependence on the condition number $\kappa_{A'}$, and not on $\kappa_A$. However, due to the fact that the difference between $S'$ and $S$ is only on the boundary nodes, this is expected to not be significant. The reasoning here is that this will depend on the norm $\|S'-S\|$, however this is a matrix of zeroes except for nodes on the boundary of the simulation region, and so will not scale with the volume of the region and should be sub-leading relative to other terms. As we shall see in Sec.~\ref{sec:condition}, this is indeed confirmed by our numerical studies.


\subsection{Driving}
\label{sec:driving}

We now proceed to extending the model and quantum algorithm from Ref.~\cite{jennings2025end-to-end} to accommodate flows driven by a divergence-free body force or a moving boundary. 


\subsubsection{Extending the model}

In LBM, one can include the effect of a divergence-free body force $\v{F}(\v{r})$ by discretizing this force in velocity space, such that its discrete representation $\phi_m(\v{r}^\star)$, ensures that the zeroth moment respects mass conservation, $\sum_m \phi_m = 0$, and that the first moment recovers the forcing applied in the momentum equation, $\sum_m \v{e}_m \phi_m = \v{F}$. While a second-order accurate discretization (in space and time) can be obtained following Ref.~\cite{silva2012first}, by adding $\phi_m$ to the right hand side of Eq.~\eqref{eq:sLBE_col_mat} and including a velocity shift, we will instead pursue a first-order accurate approach as this does not require modifications of collision matrices $F_1$ and $F_2$. As the implementation of a no-slip velocity boundary condition also includes a component which forces the flow (see Eq.~\eqref{eq:bounce_back}), we incorporate this driving which acts locally at $\v{r}^\star_b$ as:
\begin{aligns}
    g_m(\v{r}^\star_b,t) \rightarrow  g_m(\v{r}^\star_b,t) + \phi_m(\v{r}^\star_b),
\end{aligns}
such that $\phi_m(\v{r}^{*}_{b})$ in this case corresponds to $-2 w_m \frac{\v{e}^\star_m \cdot \v{u}^{*}_{w}}{ {c^{*}_s}^2 }$, the correction of the bounce back rule for a moving boundary. Our matrix formulation of the problem from Eqs.~\eqref{eq:sLBE_col_mat}-\eqref{eq:sLBE_str_mat} then gets extended by a driving step as follows:
\begin{aligns}
    \label{eq:sLBE_col_mat_new}
    \v{g}^C(t^\star) & = \left(  I+ F_1\right)\v{g}(t^\star)+ F_2 \v{g}(t^\star)^{\ot 2},\\
    \label{eq:sLBE_str_mat_new}
    \v{g}^S(t^\star) & = S\v{g}^C(t^\star), \\
    \label{eq:sLBE_dri_mat_new}
    \v{g}(t^\star+1) & = \v{g}^S(t^\star) + F_0,
\end{aligns}
where $F_0$ is the driving vector given by: 
\begin{equation}
    \label{eq:driving_vec}
    F_0 = \sum_{\v{r}^\star} \sum_{m=1}^Q  \phi_m(\v{r}^\star) \ket{\v{r}^\star} \ot \ket{m}.
\end{equation}
The effect that this third driving update rule from Eq.~\eqref{eq:sLBE_dri_mat_new} has on the Carleman vector can be inferred from its effect on $k$ copies of the state vector:
\begin{align}
    [\v{g}(t^\star+1)]^{\otimes k}&=\left[ \v{g}^S(t^\star) + F_0 \right]^{\otimes k}
    =\sum_{l=0}^k \left[I^{\otimes l} \otimes F_0^{\otimes k-l}+\mathrm{perms.}\right] \v{g}^S(t^\star)^{\otimes l} =\sum_{l=0}^{k} D_l^k \v{g}^S(t^\star)^{\otimes l},
\end{align}
where we have introduced $d^k\times d^l$ matrices:
\begin{equation}
\label{eq:Dkl}
    D_l^k:= I^{l} \otimes F_0^{\otimes k-l}+\mathrm{perms.},
\end{equation}
and where $+ \ \mathrm{perms.}$ denotes a sum over distinct permutations over $k$ subsystems, with $l$ subsystems of type 1, and $k-l$ subsystems of type 2. This means that a single time step of the evolution can be compactly written as:
\begin{align}
    \label{eq:carl_evol_d}
    \v{y}(t^\star+1) = \D\S'\C\v{y}(t^\star) + \F_0,
\end{align}
with the \emph{Carleman driving matrix}:
\begin{equation}
\D = 
\begin{pmatrix}
    I & 0 & 0      &\dots&0\\
    D^2_1     & I^{\ot 2}  & 0 &\dots&0\\
    D^3_1     & D^3_2     & I^{\ot 3}  &\dots&0\\
    \vdots&\vdots & \vdots& \ddots &\vdots\\
    D^{N_C}_1 & D^{N_C}_2 & D^{N_C}_3 & \dots & I^{\ot N_C} 
\end{pmatrix}, 
\end{equation}
and the \emph{Carleman driving vector}:
\begin{equation}
    \F_0 = [F_0, F_0^{\ot 2},\dots,F_0^{\ot N_C}]^\top.
\end{equation}

The linear system that captures the full evolution over $N_t$ time steps, including non-trivial geometry encoded by $\S'$ and driving encoded by $\D$ and $\F_0$, is now given by:
\begin{align}
    \label{eq:AH_driving}
        \begin{pmatrix}
            I& 0 & 0 & \dots&0&0\\
            -\D\S'\C   & I& 0 & \dots&0&0\\
            0     & -\D\S'\C    & I&\dots&0&0\\
            \vdots&\vdots &  \vdots& \ddots&\vdots&\vdots \\
            0&0&0&\dots&I&0\\
            0&0&0&\dots&-\D\S'\C&I
        \end{pmatrix}
        \begin{pmatrix}
            \v{y}(0)\\
            \v{y}(1)\\
            \v{y}(2)\\
            \vdots\\
            \v{y}(N_t-1)\\
            \v{y}(N_t)
        \end{pmatrix} = 
        \begin{pmatrix}
            \v{y}_{\mathrm{ini}}\\
            \F_0\\
            \F_0\\
            \vdots\\
            \F_0\\
            \F_0
        \end{pmatrix},
\end{align}
and we will again write it compactly as $A''\v{Y}=\v{b}''$.


\subsubsection{Algorithmic modifications}

The algorithmic modifications needed to include driving are as follows. First, instead of block-encoding $A'$, we now need to block-encode $A''$. And second, instead of constructing a state preparation unitary $U_{\v{b}}$ for the undriven system, we need to construct $U_{\v{b}''}$. In both cases, the central new object is the driving vector $F_0$. Using the quantum data encoding proposed in Ref.~\cite{jennings2025end-to-end}, preparing $F_0$ is equivalent to block-encoding a square matrix $\bar{F}_0$ such that its first column is $F_0$. Note that as long as $\|F_0\|\leq 1$, one can block-encode $\bar{F}_0$ with a unit prefactor using a single ancillary qubit. Moreover, the choice of LBE parameters proposed by Ref.~\cite{jennings2025end-to-end}, which guarantees that the norm of the LBE state vector is always bounded by 1, also ensures that the condition $\|F_0\|\leq 1$ is satisfied.

We start with constructing a block-encoding of $A''$. As in the case of $A'$ analyzed in the previous section, it is straightforward to show that the block-encoding parameters for $A''$ can be expressed via:
\begin{equation}
    \label{eq:A_dprime_BE_ini}
    \alpha_{A''} = \alpha_\D\alpha_{\S'} \alpha_\C + 1,\qquad n_{A} = n_\D + n_{\S'} + n_\C+2.
\end{equation}
Hence, the actual task is to block-encode the Carleman driving matrix $\D$. For that, we will use some well-known block-encoding results~\cite{lin2022lecture}. First, employing linear combination of unitaries (LCU), if a matrix $M$ can be written as:
\begin{equation}
    M = \sum_{k=1}^K M_k,
\end{equation}
where each $M_k$ can be block-encoded with a prefactor $\alpha_{M_k}$ using $n_{M_k}$ ancillary qubits, then one can construct a block-encoding of $M$ with parameters:
\begin{equation}
    \label{eq:LCU}
    \alpha_M = \sum_{k=1}^K \alpha_{M_k},\qquad n_M = \log K + \max_{k\in\{1,\dots,K\}} n_{M_k}.
\end{equation}
A variation of LCU applicable to block-diagonal matrices of the form:
\begin{equation}
    M = \sum_{k=1}^K \ketbra{k}{k} \ot M_k
\end{equation}
allows one to block-encode $M$ with parameters:
\begin{equation}    
    \label{eq:LCU_block}
    \alpha_M = \max_{k\in\{1,\dots,K\}} \alpha_{M_k},\qquad n_M = 1 + \max_{k\in\{1,\dots,K\}} n_{M_k}.
\end{equation}
Finally, for any $m\in\{1,\dots,N_C-1\}$, the following $N_C$-dimensional shift matrix,
\begin{equation}
    \Delta_m = \sum_{k=m+1}^{N_C} \ketbra{k}{k-m},
\end{equation}
can be block-encoded with a unit prefactor using a single ancillary qubit, as it is a partial permutation matrix.

Now, let us write $\D$ explicitly, using the encoding with the first register keeping track of the Carleman order:
\begin{align}
    \D = \sum_{k=1}^{N_C} \sum_{l=1}^k \ketbra{k}{l} \ot D^k_l = \sum_{k=1}^{N_C} \sum_{m=0}^{k-1} \ketbra{k}{k-m} \ot D^k_{k-m} = \sum_{m=0}^{N_C-1} \sum_{k=m+1}^{N_C} \left(\ketbra{k}{k} \ot D^k_{k-m}\right)\left(\Delta_m \ot I\right) =: \sum_{m=0}^{N_C-1} B_m.
\end{align}
Using Eq.~\eqref{eq:LCU}, we then have:
\begin{equation}
    \alpha_\D = \sum_{m=0}^{N_C-1} \alpha_{B_m},\qquad n_\D = \log N_C + \max_{m\in\{0,\dots,N_C-1\}} n_{B_m},
\end{equation}
and the problem reduces to constructing block-encodings of $B_m$ matrices. To achieve this, we can block-encode the two factors of $B_m$ separately. As already mentioned, the second factor, $\Delta^m\ot I$, can be block encoded with a unit prefactor and one ancillary qubit, whereas for the first factor, we will employ Eq.~\eqref{eq:LCU_block}. This results in the following block-encoding parameters for $B_m$:
\begin{equation}
    \alpha_{B_m} = \max_{k\in\{m+1,\dots,N_C\}} \alpha_{D^k_{k-m}},\qquad n_{B_m} = 2+ \max_{k\in\{m+1,\dots,N_C\}} n_{D^k_{k-m}}.
\end{equation}
Finally, recalling Eq.~\eqref{eq:Dkl}, we note that each $D^k_{k-m}$ consists of a sum over $\binom{k}{m}$ terms, each containing $m$ copies of $\bar{F}_0$. Hence, using Eq.~\eqref{eq:LCU}, we have:
\begin{equation}
    \alpha_{D^k_{k-m}} = \binom{k}{m},\qquad n_{D^k_{k-m}}=\log\binom{k}{m} + m.
\end{equation}
Putting it all together, we arrive at:
\begin{aligns}
    \alpha_\D=&\sum_{m=0}^{N_C-1} \binom{N_C}{m} = 2^{N_C}-1,\\ 
    n_\D =& \log N_C + 2+\max_{m\in\{0,\dots,N_C-1\}} \left(\log\binom{N_C}{m} + m \right)\leq \log N_C + 2 N_C+1.
\end{aligns}

Substituting the expressions derived above, together with the previous results from Eq.~\eqref{eq:A_prime_BE}, into Eq.~\eqref{eq:A_dprime_BE_ini}, we thus conclude that:
\begin{equation}
    \alpha_{A''} = (2^{N_C}-1)2^{N_C} \alpha_\C + 1 \approx 4^{N_C} \alpha_A,\qquad n_{A''} = n_A + 5N_C + \log N_C +1.
\end{equation}
Thus, the complexity cost related to extending the algorithm from Ref.~\cite{jennings2025end-to-end} to include both non-trivial geometries and forcing increases the query complexity by a factor of $4^{N_C}$, and requires additional $5N_C$ ancillary qubits (up to leading order in $N_C$). As already discussed in the previous section, this cost should not strongly affect the algorithm's efficiency, as $N_C$ can be chosen to be small in the sub-threshold regime, due to exponential convergence of the Carleman expansion. However, there are two other sources of potential complexity increase: the difference between the condition numbers of $A$ and $A''$, and the cost related to state preparation unitary $U_{\v{b}''}$ instead of $U_{\v{b}}$. We postpone the discussion of the first effect to Sec.~\ref{sec:condition}, and now focus on constructing $U_{\v{b}''}$.

Let us first assume that one can efficiently construct a unitary $U_{F_0}$ that prepares a quantum state $\ket{F_0}$ that is a normalized version of the driving vector $F_0$ from Eq.~\eqref{eq:driving_vec}. Then, as was shown in Ref.~\cite{liu2021efficient}, one can prepare $\ket{\F_0}$ using $O(N_C)$ queries to $U_{F_0}$ and $O(N_C)$ single qubit gates. Hence, each query to a state preparation unitary $U_{\F_0}$ translates into $O(N_C)$ queries to $U_{F_0}$. Next, for simplicity, let us assume that one initializes each node of the fluid system in an equilibrium state corresponding to zero velocity and zero pressure (as is typical in CFD simulations, and as we do in all examples presented in Sec.~\ref{sec:case_studies}). This then means that $\v{y}_{\mathrm{ini}}$ is a zero vector. Additionally, adding a single driving step before the start of the evolution, we can rewrite $\v{b}''$ as
\begin{equation}
    \v{b}'' = \sum_{t^\star=0}^{N_t} \ket{t^\star} \ot \F_0.
\end{equation}
Hence, preparing $\ket{\v{b}''}$ just requires a uniform state preparation on the time step register and a single preparation of $\ket{\F_0}$, which in turn requires $O(N_C)$ uses of $U_{F_0}$. Therefore, one can conclude that, as long as one can efficiently prepare $\ket{F_0}$, one can also efficiently prepare $\ket{\v{b}''}$.

Although the complexity cost of preparing $\ket{F_0}$ may be significant for random or unstructured driving, in practice one is usually interested in driving with explicit functional dependence on position. For example, in the lid-driven cavity flow and flow past a cylinder, analyzed in the next section, driving has a particularly simple form:
\begin{equation}
    \label{eq:driving_vec_simple}
    F_0 = \sum_{\v{r}^\star\in\mathcal{R}}\ket{\v{r}^\star} \ot \sum_{m=1}^Q\phi_m \ket{m}.
\end{equation}
In the above, $\mathcal{R}$ is a subset of nodes (neighboring the lid and inlet ones, respectively) and $\phi_m$ are just a constant driving coefficients encoded in a small velocity register (consisting of $2D$ qubits for fluid in $D$ dimensions). Hence, we just need a uniform state preparation on the position register and a general state preparation on a small velocity register. The complexity cost of preparing $\ket{F_0}$ is thus constant. Therefore, we will assume that the complexity cost of $U_{\v{b}''}$ is negligible compared to the cost of unitary block-encoding $U_{A''}$ of the linear system.


\section{Numerical studies}
\label{sec:case_studies}

In Sec.~\ref{sec:framework}, we introduced a quantum algorithm based on a Carleman embedding of the incompressible lattice Boltzmann method as described in Ref.~\cite{jennings2025end-to-end}. In the previous Section~\ref{sec:extensions}, we demonstrated how this approach can be extended to model non-trivial flows in domains which do not have periodic boundary conditions. In particular, we showed how: no-slip boundary conditions, free-slip boundary conditions, body forcing and non-trivial geometries can be accommodated. In this section, we implement and numerically test these theoretical developments in order to assess the ability of the Carleman embedding to approximate three different flows demanding these extensions.

We focus on the driven Taylor-Green vortex, lid-driven cavity flow, and the flow past a cylinder, enclosed within a simulation region of size $L_x \times L_y$. They are described by the incompressible Navier-Stokes equations:
\begin{subequations}
    \begin{align}
        \partial_t \v{u} + \left( \v{u} \cdot \nabla \right) \v{u} &= - \frac{1}{\rho_0}\nabla p + \nu \nabla^2 \v{u},
        \label{eq:Momentum}\\
        \nabla \cdot \v{u} &= 0,
        \label{eq:divU}
    \end{align}
\end{subequations}
together with an appropriate set of boundary conditions. In each case, the forcing applied (via the boundary conditions or via a body forcing $\v{F}$), has an associated characteristic velocity-scale $U$ and length-scale $L$, which allows a physical Reynolds number $\re=UL/\nu$ to be constructed. For the lid-driven cavity flow, for example, these scales correspond to the speed of the wall and the cavity size, while for the flow past a cylinder they correspond to the inlet velocity and the cylinder diameter. However, given that we numerically solve the lattice Boltzmann equation, Eq.~\eqref{eq:LBE_incomp}, which is non-dimensionalized in terms of lattice units (i.e., using a velocity scale $\Delta x/\Delta t$ and length-scale $\Delta x$), it is useful to first establish the relationship between $\re$ and these lattice units. As described in Ref.~\cite{jennings2025end-to-end}, guided by turbulence theory and the scaling required to maintain $\|\v{g}\| \sim O(1)$, we set:
\begin{equation}
    \ma \approxeq \frac{u^{\star}_0}{c_s^\star\sqrt{N_x N_y}},
    \label{eq:Ma_scaling}
\end{equation}
to obtain this relation. The dimensionless parameter $u^{\star}_0$ introduced as a control of the temporal resolution is set to $u^{\star}_0=1$ unless otherwise stated. For a periodic box, the number of grid-points per direction required to resolve homogeneous isotropic turbulence scales as $N_x \gtrsim \eta_x \re^{3/4}$ and $N_y \gtrsim \eta_y \re^{3/4}$, where $\eta_i=L_i/L$ (see Ref.~\cite{pope2001turbulent}). On this basis, we fix $N_i = \eta_i Re^{\beta}$ with $\beta \geq 3/4$, while for the temporal resolution we make use of the advection time-scale $T = L/U$, which non-dimensionally translates into the number of time-steps $N_t = \frac{1}{\Delta t} \frac{L}{U}$. Using the definition of the Mach number $\ma = U / (c_s^\star \frac{\Delta x}{\Delta t})$, together with Eq.~\eqref{eq:Ma_scaling}, we can eliminate $\Delta t$ giving $N_t = N_xN_y/( u^\star_0\sqrt{\eta_x\eta_y})$. For a given simulation with fixed $\eta_x$ and $\eta_y$, it therefore suffices to specify $\re$ and $\beta$, as this fixes the remaining parameters according to:
\begin{equation}
    \label{eq:parameter_choice}
    N_x = \left\lceil \eta_x\re^\beta \right\rceil ,\qquad N_y = \left\lceil \eta_y\re^\beta \right\rceil ,\qquad N_t = \left\lceil N_xN_y/( u^\star_0 \sqrt{\eta_x\eta_y}) \right\rceil,\quad \tau^\star = \frac{3u_0^\star}{\sqrt{\eta_x\eta_y}}\frac{1}{\re}.
\end{equation}
Varying $\beta$ and $u^\star_0$ then allows one to control the spatial and temporal resolution if needed.

Our numerical implementation of the classical LBM and Carleman LBM is programmed in such a way that linear operations including streaming (which incorporates the boundary conditions) and forcing are represented as matrices. These matrices are then used by both our classical and Carleman LBM routines when performing these operations. This construction was intentionally chosen, as by validating our classical LBM code, we in-turn validate all linear components of the Carleman LBM. We performed validation tests by correctly reproducing the time evolution of the decaying and forced Taylor-Green vortex with second-order error in both space $O(\Delta x^2)$ and time $O(\Delta t^2)$ (see Fig.~\ref{fig:TG_Convergence_Validation} in the Appendix). Given that the Taylor-Green vortex does not test the validity of our implementation of the no-slip and slip-free boundary conditions, we also performed our simulations with a second independent code, which we used to verify our implementation. 

To assess the ability of the Carleman LBM to correctly reproduce the classical LBM, we use the relative $L_2$ error:
\begin{equation}
    \epsilon_{\rel}(t^\star) := \frac{\| \v{u}^{N_C}(\v{r}^\star,t^\star) - \v{u}(\v{r}^\star,t^\star) \|}{ \| \v{u}(\v{r}^\star,t^\star) \|}, \qquad \epsilon_{C} := \underset{t^\star \in \{1, \cdots, N_t\}}{\text{max}} \epsilon_{\rel}(t^\star),
    \label{eq:Carleman_error}
\end{equation}
where $\| \v{u}(\v{r}^\star,t^\star) \|^2 = 1/V \int |\v{u}|^2 dV$ and superscript $N_C$ indicates the Carleman LBM solution with truncation order~$N_C$. Note that only $t^\star>0$ is considered, as any error at $t^\star=0$ is independent of the embedding method. Although it would, in principle, be desirable to fix $\re \gg 1$ and investigate how $\epsilon_C$ depends on the Carleman truncation order $N_C$, the dimension of the linear embedding from Eq.~\eqref{eq:carl_evol_d} (as illustrated by Table~\ref{table:dimensions}) grows with $N_C$ according to Eq.~\eqref{eq:Carleman_dimension}, thus making this calculation extremely demanding for classical computers. Nevertheless, in order to at least attempt to accurately simulate physically relevant configurations, we must choose sufficiently large $N_x = \eta_x\re^{\beta}$, which in turn restricts our calculations to $N_C=1,2,3$. Although this is compounded by the fact our code is currently limited to serial runs, we envisage making use of GPUs in future studies.
\begin{table}[t]
\centering
\begin{tabular}{|c|c|c|c|c|c|c|}
    \hline
    \multirow{2}{*}{$\re$} 
      & \multicolumn{2}{c|}{$N_C=1$} 
      & \multicolumn{2}{c|}{$N_C=2$} 
      & \multicolumn{2}{c|}{$N_C=3$} \\
    \cline{2-7}
      & $d_C$ & $d_C(N_t+1)$ 
      & $d_C$ & $d_C(N_t+1)$ 
      & $d_C$ & $d_C(N_t+1)$ \\
    \hline
    10   & $9\cdot 10^2$      & $9\cdot 10^4$ 
         & $8.1\cdot 10^5$    & $8.1\cdot 10^7$ 
         & $7.3\cdot 10^{8}$  & $7.3\cdot 10^{10}$ \\ 
    100  & $9\cdot 10^4$      & $9\cdot 10^8$ 
         & $8.1\cdot 10^9$    & $8.1\cdot 10^{13}$ 
         & $7.3\cdot 10^{14}$ & $7.3\cdot 10^{18}$ \\ 
    1000 & $9\cdot 10^6$      & $9\cdot 10^{12}$ 
         & $8.1\cdot 10^{13}$ & $8.1\cdot 10^{19}$ 
         & $7.3\cdot 10^{20}$ & $7.3\cdot 10^{26}$ \\ 
    \hline
\end{tabular}
\caption{Dimensions $d_C$ of the Carleman vector and $d_C(N_t+1)$ of the Carleman linear system for two-dimensional LBE problems with resolution parameter $\beta=1$, different Reynolds numbers $\re$ and truncation orders $N_C$.}
\label{table:dimensions}
\end{table}

Having detailed our numerical implementation, we now  consider our three examples: how $\epsilon_C$ depends on $\re$ and~$N_C$, how the spatial form of the absolute value of the difference $| \v{u}^{N_C}(\v{r}^\star,t^\star) - \v{u}(\v{r}^\star,t^\star) |$ varies with $N_C$, and how $\epsilon_{\rel}(t^\star)$ varies in time. We consider the driven Taylor-Green vortex first as it is the simplest case, followed by the lid-driven cavity, and the flow past a cylinder. Finally, we will present numerical results on the condition number of the associated linear systems, thus addressing the question about the complexity cost of simulating non-trivial flows.


\subsection{Driven Taylor-Green vortex}
\label{sec:vortex}

In a 2D domain $\Omega := [0,L] \times [0,L]$ with periodic boundary conditions, the Taylor-Green vortex state is given by:
\begin{equation}
    \v{u}(\v{r},t) = [ \sin ( k x) \cos (k y), -\cos (k x) \sin (k y) ] f(t), \quad p(\v{r},t)    = -\frac{\rho_0}{2}(\sin^2 (kx) + \sin ^2 (ky) ) f^2(t),
    \label{eq:Taylor_Green}
\end{equation}
where $\v{r} = (x,y)$, the wavenumber $k = 2 \pi / L$, and $f(t)$ is explicit time dependence. Substituted into the Navier-Stokes equations, Eqs.~\eqref{eq:Momentum}-\eqref{eq:divU}, this state leads to the cancellation of the pressure gradient and non-linear terms, leaving $\partial_t \v{u} = \nu \Delta \v{u}$, which when solved with the initial condition $f(t=0) = U_{TG}$ gives $f(t) = U_{TG} \exp(-2 \nu k^2 t)$. As this solution decays exponentially in time, it is interesting to determine the form of a forcing $\bm{F}$, which when added to the right hand side the Navier-Stokes equations not only sustains Eq.~\eqref{eq:Taylor_Green}, but also leaves open the possibility of producing time-dependent states as the forcing increases in strength. Setting $\partial_{t} \v{u}= 0$ and substituting Eq.~\eqref{eq:Taylor_Green} into these equations implies a forcing of the form:
\begin{equation}
    \v{F} = 2 \nu k^2 \v{u}(\v{r},t=0).
    \label{eq:Taylor_Green_forcing}
\end{equation}
Choosing the velocity scale $U_{TG}$ and length-scale $L$, we can non-dimensionalize the problem such that $\re=U_{TG}L/\nu$ characterizes the relative strength of this forcing. In order to test the extension of the Carleman LBM to include driving as described in Sec.~\ref{sec:driving}, we follow the approach of Ref.~\cite{silva2012first} that considered the addition of a divergence-free body force for the incompressible LBM~\cite{he1997lattice}. For a steady or slowly varying flow such that $\partial_t \v{u} \approxeq 0$, it is consistent to add
\begin{equation}
    \label{eq:vor_driving}
    \phi_m = w_m \frac{\v{e}_m \cdot \v{F}}{c_s^2}
\end{equation}
to the right hand side of Eq.~\eqref{eq:LBE_incomp}. As Eq.~\eqref{eq:Taylor_Green} is a steady solution of the forced problem, we validated our implementation of the driving is \emph{nearly} second-order accurate\footnote{The velocity shift proposed in Ref.~\cite{silva2012first} requires modifications of $F_1$ and $F_2$ which are not currently included in our code.} as shown in Fig.~\ref{fig:TG_Convergence_Validation} in the Appendix. However, in order to conduct a more strenuous test of the Carleman convergence, we choose instead to investigate transient dynamics, by starting from the zero initial condition $\v{u}(\v{r},t=0)=0, \; p(\v{r},t=0) = 0$ and tracking how the classical and Carleman LBM evolve to the steady state solution from Eq.~\eqref{eq:Taylor_Green}. 

\begin{figure}[t!]
    \centering
    \includegraphics[width=0.49\linewidth]{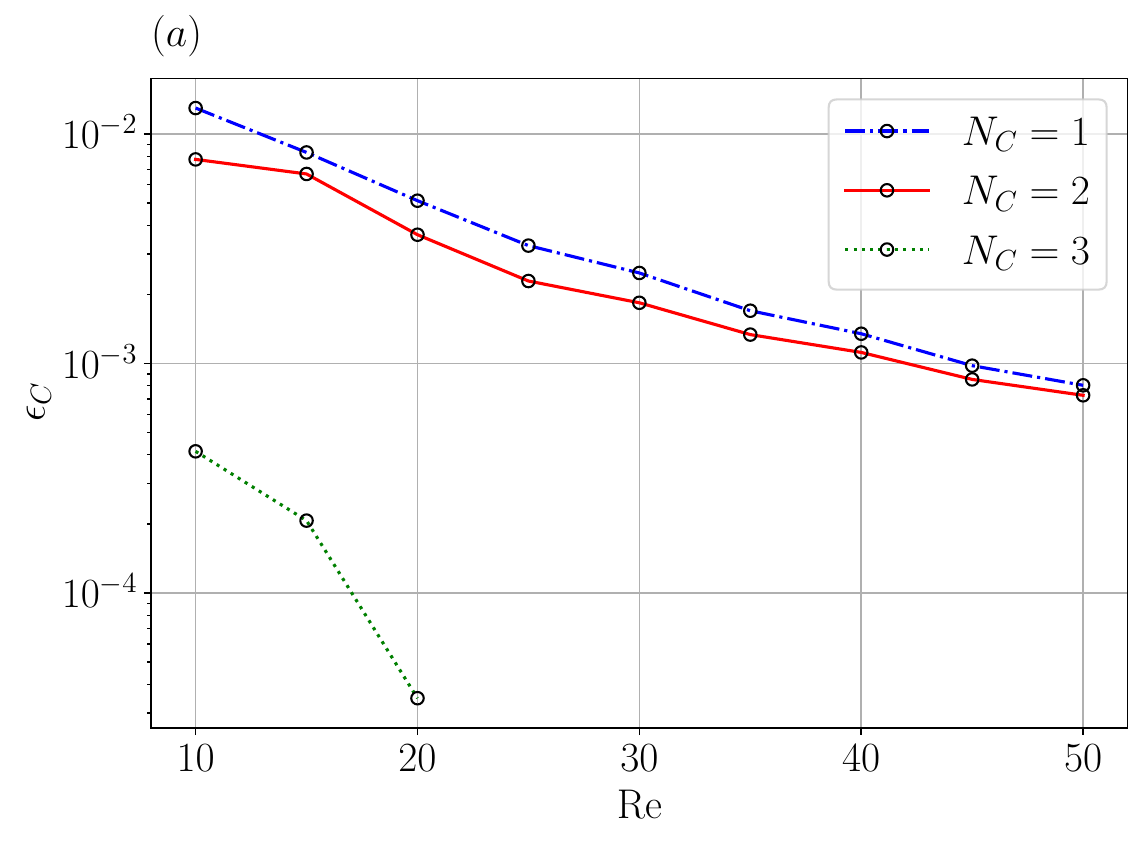}
    \includegraphics[width=0.49\linewidth]{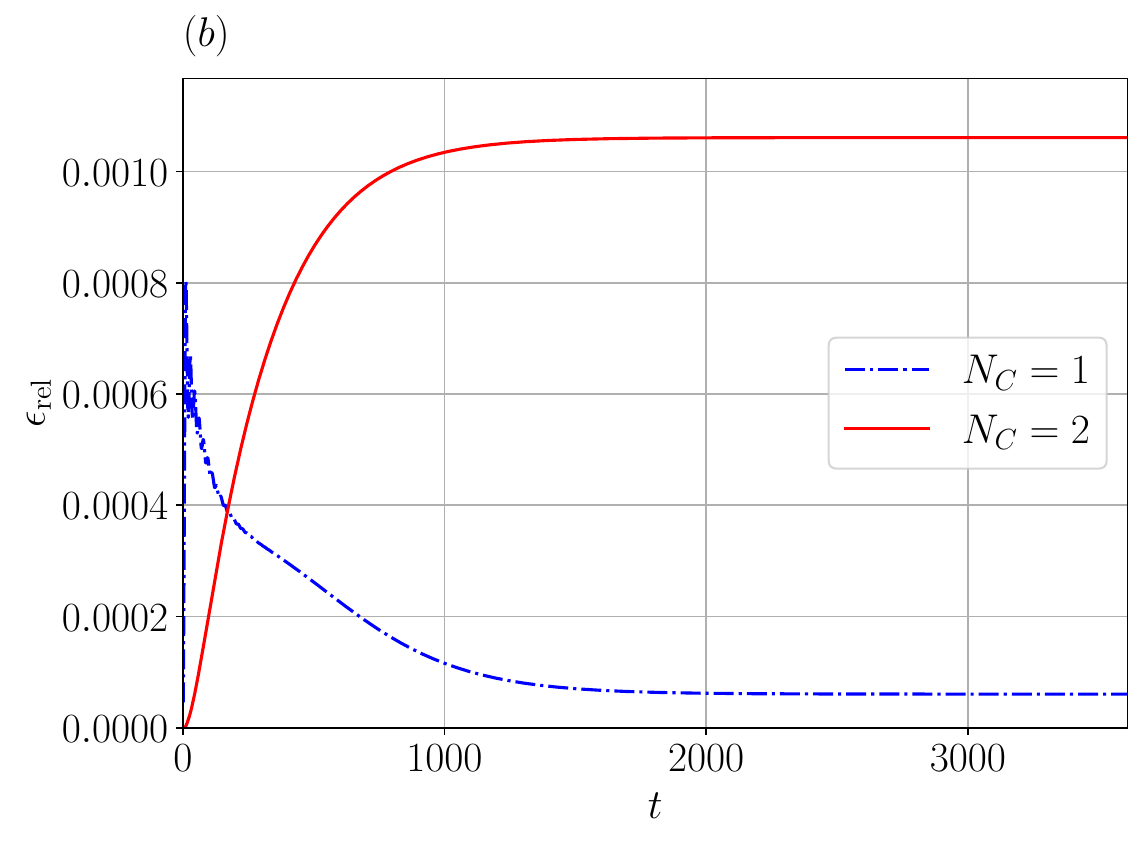}
    \caption{Carleman truncation error for the forced Taylor-Green vortex flow with $\beta=0.75$. Left: Dependence of~$\epsilon_C$ on the Reynolds number~$\re$ and truncation order $N_C=1,2,3$ for one advection time. Right: Time evolution of $\epsilon_{\rel}$ over $10$ advection times for $N_C=1,2$ and $\re=50$.}
    \label{fig:TGF_error}
\end{figure}

Figure~\ref{fig:TGF_error}(a) shows the convergence behavior of the Carleman error $\epsilon_C$ for $N_C=1,2,3$ with $\beta=0.75$. We observe that the error for the linear Carlemen LBM, $N_C=1$, decreases with increasing $\re$. This was also observed for the decaying Taylor-Green vortex~\cite{jennings2025end-to-end}, and may be attributed to the fact that as $N_x = \re^{\beta}$ increases, the Mach number reduces in accordance with Eq.~\eqref{eq:Ma_scaling}, thereby reducing the strength of nonlinearity. Although the quadratic Carleman expansion does not significantly improve the relative error, by continuing to $N_C=3$ we do find a dramatic reduction. This demonstrates that at least for small $\re$ the proposed Carleman LBM for forced flows can converge as $N_C$ increases, albeit not necessarily uniformly. Due to the need to consider $N_C=3$ to observe a convincing trend towards convergence, it was necessary to limit our spatial resolution to $\beta=0.75$. However, by recomputing Fig.~\ref{fig:TGF_error}(a) for $\beta=1$ (without $N_C=3$ included), we verified that the results presented are consistent with those observed for a larger resolution. 

\begin{figure}[t!]
    \centering
    \includegraphics[width=\linewidth]{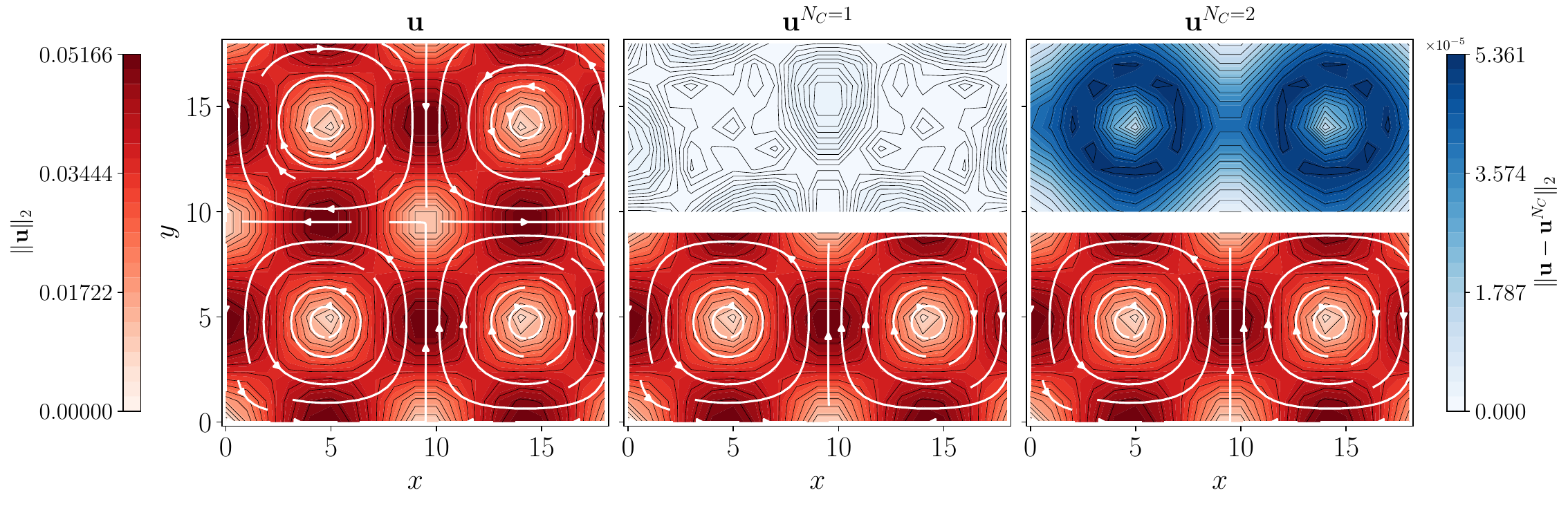}
    \caption{Spatial form of the classical LBM solution, the Carleman LBM solution and their associated error for the forced Taylor-Green vortex flow after $10$ advection times for $\beta=0.75$ and $\re=50$. Left: Classical LBM solution. Middle: Linear Carleman approximation (bottom) and error (top). Right: Quadratic Carleman approximation (bottom) and error (top).}
    \label{fig:TGF_field_error}
\end{figure}

To assess the time evolution of the truncation error, we chose $\re=50$ (corresponding to $N_x=N_y=19$) and time-integrated the zero initial condition subject to the Taylor-Green forcing for $10$ advection times. As shown in Fig.~\ref{fig:TGF_error}(b), the relative error $\epsilon_{\rel}$ plateaus following an initial transient. This shows that error associated with the Carleman LBM for forced flows does not continually grow, but converges to a finite value. The spatial form of the solution and its associated Carleman truncation error are shown in Fig.~\ref{fig:TGF_field_error}. Although it is interesting to go to larger $\re$, we intentionally limit the study to $\re = 50$, as beyond this threshold the solution state shown in Fig.~\ref{fig:TGF_field_error} becomes unstable and can transition to another more complex steady state.  Since it is known that a single, global linearization scheme cannot describe multiple basins of attraction, it follows that a direct Carleman embedding will fail for Reynolds number above a certain value. However, recent work has begun to go beyond such obstacles by constructing adaptive linearization schemes~\cite{novikau2025globalizing}. It would be of interest to extend such methods to quantum algorithms, and the above driven Taylor-Green vortex could be an excellent test-bed for such methods.

\subsection{Lid-driven cavity}
\label{sec:lid}

To test the extension of the Carleman LBM for no-slip boundary conditions as described in Sec.~\ref{sec:extensions}, we consider lid-driven cavity flow, initially studied for the classical incompressible LBM in Ref.~\cite{guo2000lattice}. This problem considers a 2D cavity, for example a notch of size $\Omega := [0,H] \times [0,H]$ taken out of an aeroplane wing, such that the air flowing over the top surface moves at constant velocity $U_w$, causing the sheared fluid within the cavity to circulate. Letting $\v{u} = (u,v)$, we set the components of the velocity vector at the top $y=H$ and bottom $y=0$ boundaries of the cavity to
\begin{equation}
    u(x,y=H,t) = U_w, \quad v(x,y=H,t) = 0, \quad  u(x,y=0,t) = 0, \quad v(x,y=0,t) = 0,
\end{equation}
and at the sides to
\begin{equation}
    \v{u}(x=0,y,t) = \v{u}(x=H,y,t) = 0.
\end{equation}
As we use the lattice Boltzmann method, these macroscopic boundary conditions are converted into rules for how the populations $g_m$ are updated at the walls as described in Secs.~\ref{sec:Geometry}-\ref{sec:driving}. By choosing the velocity scale $U_w$ and the box size as a length-scale $H$, we can non-dimensionalize the problem such that $\re = U_w H/\nu$ characterizes the strength at which the moving top boundary drives the flow. As per the Taylor-Green vortex, we start from the zero initial condition $\v{u}(\v{r},t=0)=0, \; p(\v{r},t=0) = 0$ and track how the classical and Carleman LBM evolve to a steady state solution.

\begin{figure}[t!]
    \centering
    \includegraphics[width=0.49\linewidth]{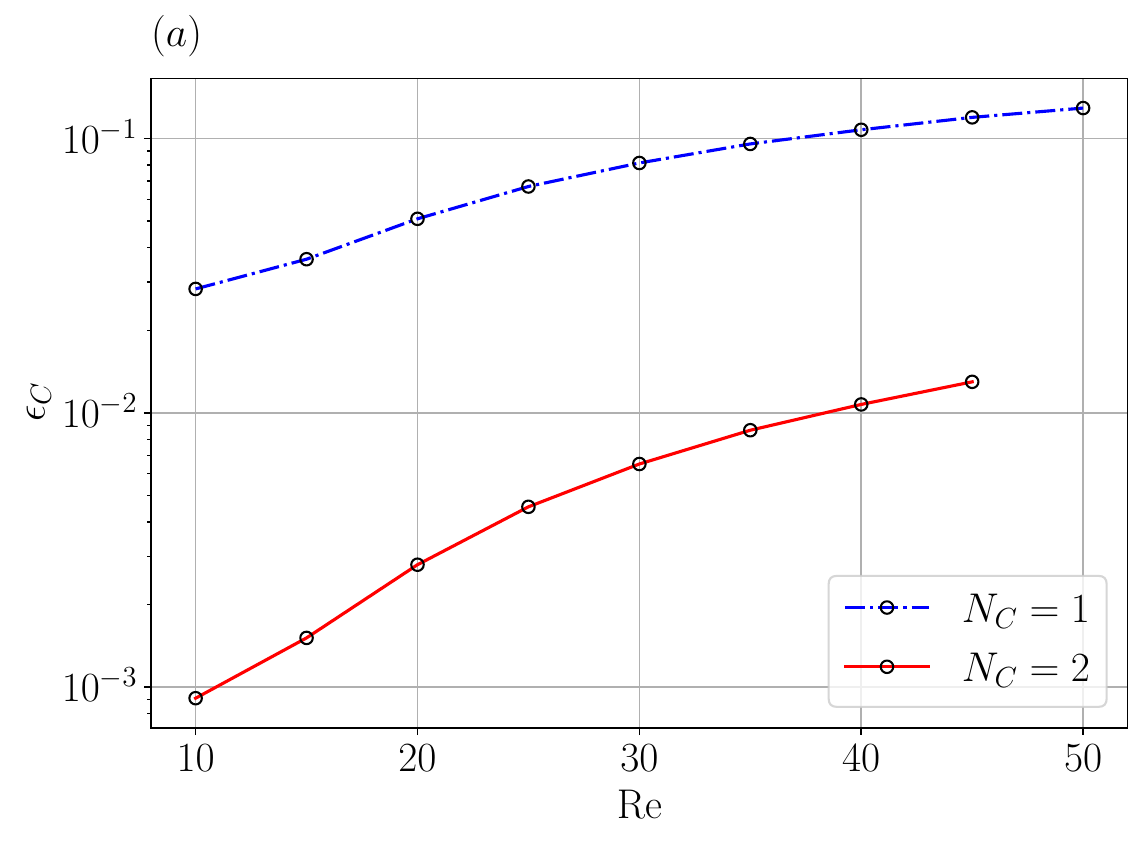}
    \includegraphics[width=0.49\linewidth]{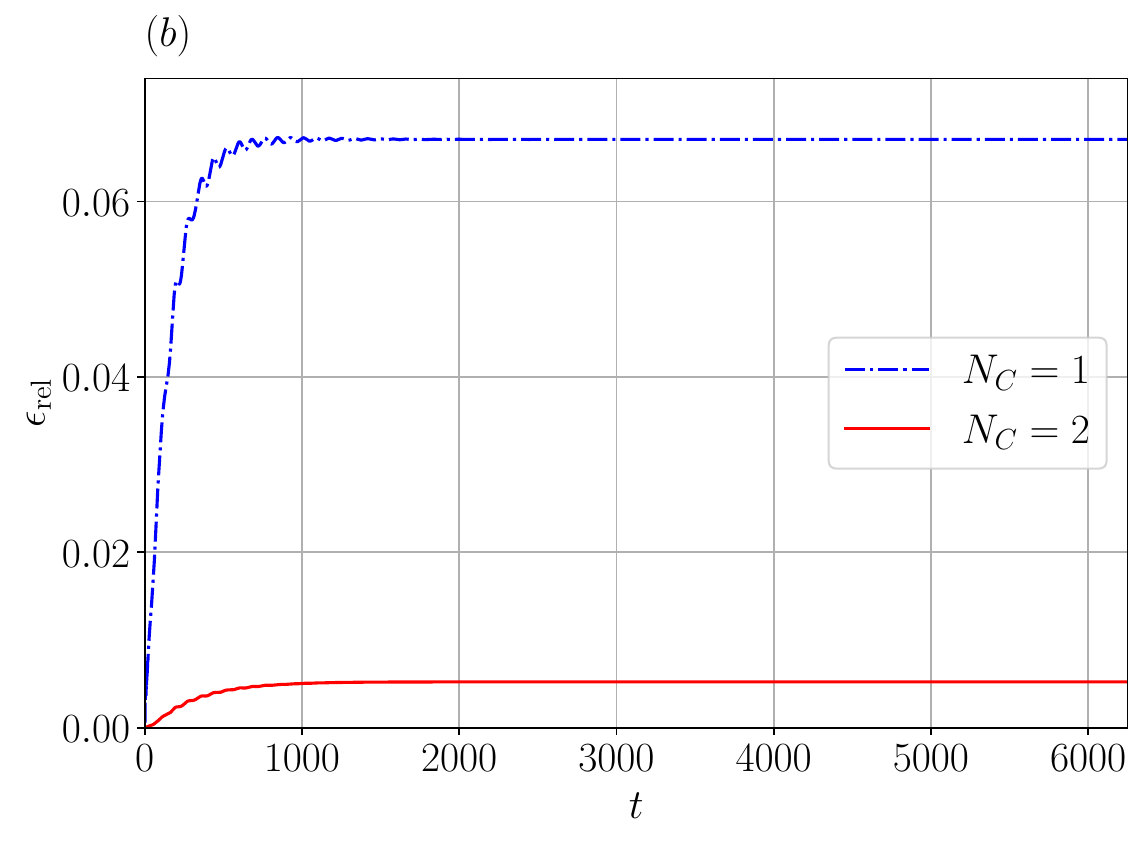}
    \caption{Carleman truncation error for the lid-driven cavity flow with $\beta=1$. Left: Dependence of $\epsilon_C$ on the Reynolds number $\re$ and truncation order $N_C=1,2$ for one advection time. Right: Time evolution of $\epsilon_{\rel}$ over $10$ advection times for $N_C=1,2$ and $\re=25$.}
    \label{fig:lid_error}
\end{figure}

\begin{figure}[t!]
    \centering
    \includegraphics[width=\linewidth]{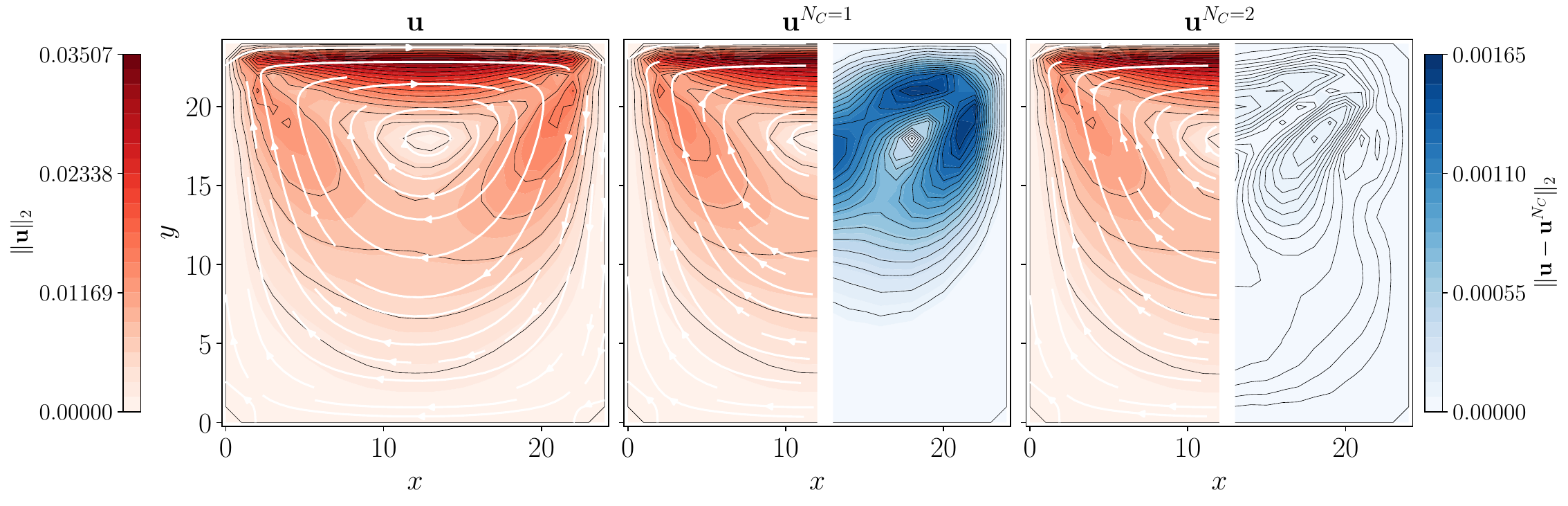}
    \caption{Spatial form of the classical LBM and Carleman LBM solutions (red) and their associated error (blue) for the lid-driven cavity flow after $10$ advection times for $\beta=1$ and $\re=25$. Left: the classical LBM solution. Middle: linear Carleman approximation. Right: the quadratic Carleman approximation.}
    \label{fig:Lid_driven_field_error}
\end{figure}

Figure \ref{fig:lid_error}(a) shows the convergence behavior of the Carleman error for $N_C=1,2$. In contrast to the previous example, we observe that as $\re$ increases, the error grows for both $N_C=1$ and $N_C=2$. Also, by going to $N_C=2$, we directly observe a big drop in the error $\epsilon_C$ which appears to shrink as $\re$ increases. On this basis, we did not go to $N_C=3$, but instead increased the spatial resolution for this example by setting $\beta=1$, such that $N_x=N_y=\re$. We attribute the increase in $\epsilon_C$ with $\re$ to the fact that the form of the solution in this case changes as $\re$ increases, and in particular that strong velocity gradients develop near the lid as shown in Fig.~\ref{fig:Lid_driven_field_error}. For the Taylor-Green vortex, by contrast, the Carleman embedding only needed to approximate the same form of solution (given by Eq.~\eqref{eq:Taylor_Green}) for all $\re$. This is because for the range of $\re$ considered, only the time-scale at which we reach the steady state changes with $\re$. 

Figure~\ref{fig:lid_error}(b) shows the time evolution of the relative error for $N_C=1,2$ at $\re=25$. Once again, we observe that the Carleman error does not continually increase in time but instead converges to a finite value. A second interesting observation is that oscillations appear in $\epsilon_{\rel}$ for both $N_C=1$ and $N_C=2$, which decay as $t$ increases. However, the amplitude of these oscillation is much larger for $N_C=1$. Similar behavior was also observed for the Taylor-Green vortex in Fig.~\ref{fig:TGF_error}(b). Taken together, the results from Fig.~\ref{fig:lid_error} demonstrate that at least for small $\re$ the proposed Carleman LBM for shear flows with no-slip boundary conditions converges as $N_C$ increases.

Figure~\ref{fig:Lid_driven_field_error} shows the spatial form of the solution predicted by the classical LBM as well as the linear $\v{u}^{N_C=1}$ and quadratic $\v{u}^{N_C=2}$ Carleman LBM approximations. Due to horizontal movement of the lid, we observe larger gradients of the velocity near the moving lid. As the strongest gradients develop near the top right hand corner, the location where the fluid traveling rightwards meets a solid boundary and is forced to re-circulate, we have chosen to highlight the form of spatial error in this region. Comparing the form of the error for the linear $\v{u}^{N_C=1}$ and quadratic $\v{u}^{N_C=2}$ approximations, we observe that $N_C=2$ further reduces the intensity of the error and shifts it away from the walls. In both solutions, we also observe that exactly at the boundaries the error is smallest. This suggests that the proposed extension of the Carleman LBM for no-slip boundary conditions yields a physically consistent approximation.

\subsection{Flow past a cylinder}
\label{sec:cylinder}

In order to computationally model an object in a wind tunnel, it is necessary to be able to accommodate inlet/outlet flows. A simple example in which the flow enters the domain, flows around the object embedded within, and then exits the domain, is the flow around a cylinder. Following Ref.~\cite{guo2000lattice} that studied this flow for the classical incompressible LBM, we use this example to test the extension of the Carleman LBM described in Section~\ref{sec:extensions} to include free-slip outflow boundary conditions, as well as no-slip boundary conditions on a cylinder embedded within the domain.

The computational domain considered is $\Omega := [-2.5D,11.5D] \times [-3.5D, 3.5D]$ where $D$ is the cylinder diameter. The cylinder is positioned at $\boldsymbol{r}=(0,0)$, slightly downstream of the inlet, but sufficiently far from the outlet, so that the influence of the outlet boundary condition is not felt upstream. In the vertical, we use periodic boundaries, at the inlet we impose the no-slip boundary condition
\begin{equation}
    \v{u} = (U_{\infty},0),
\end{equation}
according to Eq.~\eqref{eq:bounce_back}, while at the outlet we impose
\begin{equation}
    \partial_x u = \partial_x v = 0,
\end{equation}
such that the flow is free to exit the right hand side of the domain with a vertically non-uniform velocity profile. To convert the latter boundary condition into rules for how the populations $g_m$ should be updated at the outlet $\v{r}_o$, we use non-equilibrium extrapolation scheme to implement a zero-gradient (pressure and velocity) outlet\footnote{We acknowledge that a fixed pressure outlet with zero-gradient velocity handles reflections due to the finite domain better and has improved stability. Future work will consider this modification of the outflow boundary condition.}{ following Refs.~\cite{junk2008outflow, guo2013lattice}}, whose implementation is described in Sec.~\ref{sec:Geometry}. Due to the large domain required to minimize the influence of the outflow boundary on the flow, we are limited to small $\re \leq 10$ and $N_C=1,2$ with $\beta=0.75$, where the Reynolds number is defined by $\re = D U_\infty/\nu$. As before, we initialize the systems at zero initial condition with $\v{u}(\v{r},t=0)=(0,0), \; p(\v{r},t=0) = 0$.

\begin{figure}[t!]
    \centering
    \includegraphics[width=0.49\linewidth]{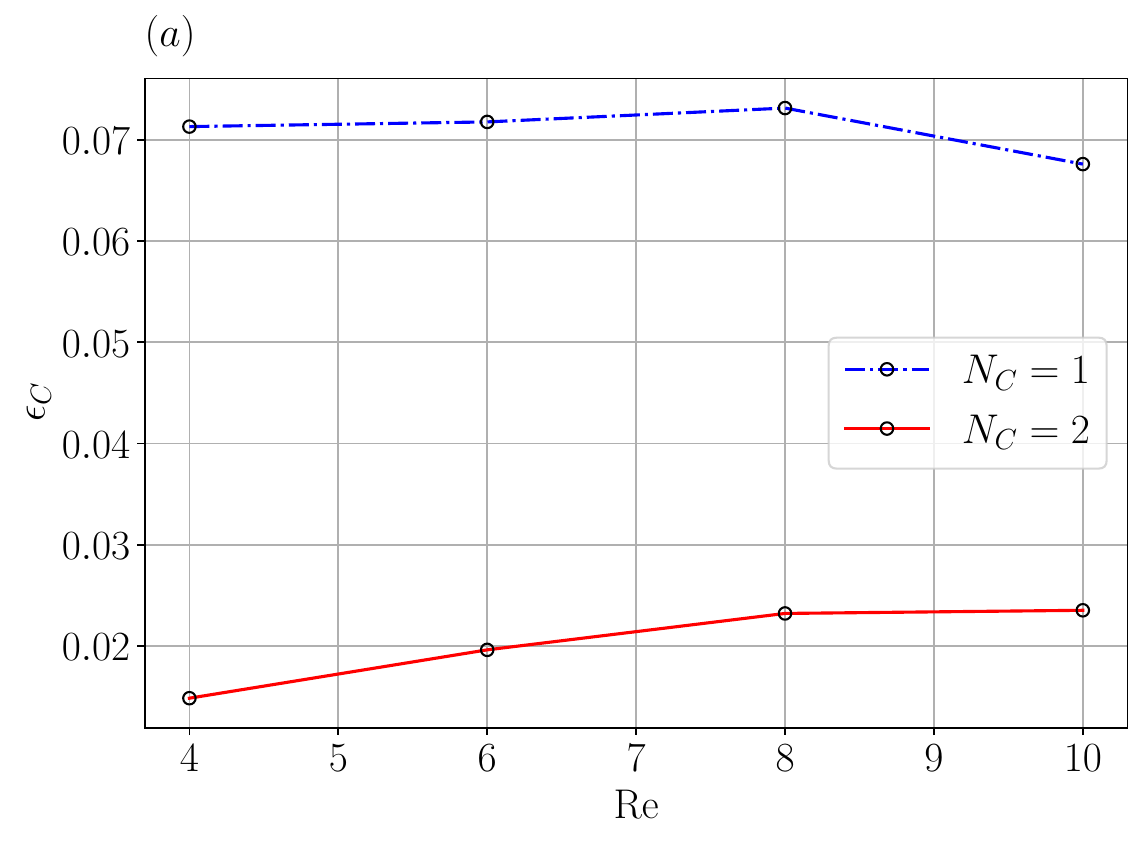}
    \includegraphics[width=0.49\linewidth]{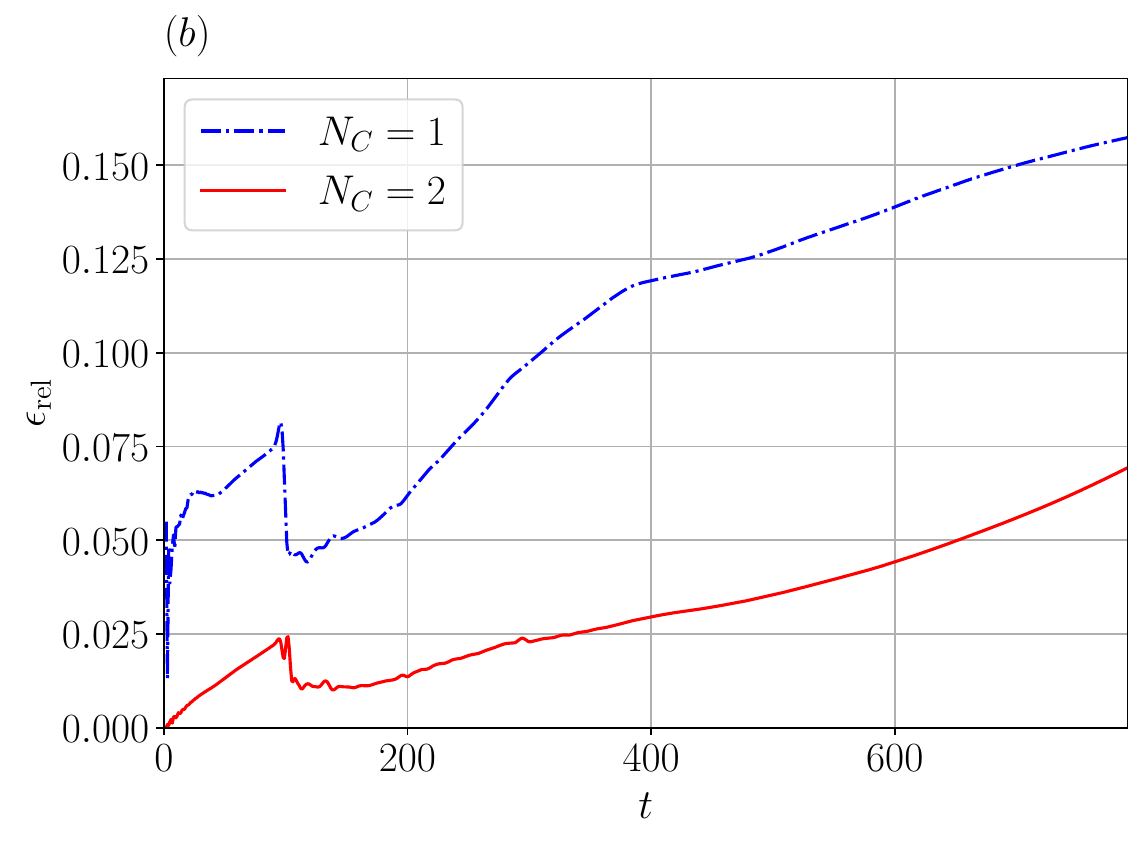}
    \caption{Carleman truncation error for the flow past a cylinder with $\beta=0.75$. Left: Dependence of $\epsilon_C$ on the Reynolds number $\re$ and truncation order $N_C=1,2$ for one advection time. Right: Time evolution of $\epsilon_{\rel}$  over $5$ advection times for $N_C=1,2$ and $\re=6$.}
    \label{fig:cylinder_error}
\end{figure}

Figure~\ref{fig:cylinder_error}(a) shows the Carleman truncation error for $N_C=1,2$ as a function of $\re$. Due to the small variations in spatial resolution with $\re$, we do not observe large differences in $\epsilon_C$ for this case. Consistent with the previous two cases considered, however, we do find that by increasing the Carleman truncation order from $N_C=1$ to $N_C=2$ we can reduce the maximum relative error over one advection time from $7\%$ to $2.5\%$. Consistent with the lid-driven cavity flow, we observe that over 5 advection times the relative error for $N_C=2$ remains less than $N_C=1$. The dynamics of the relative error for this case is, however, more complex. In particular, at $t\approxeq 100$ we observe a drop in the relative error after an initial increase. We attribute this signature of the error signal to be the time taken for the wake of the cylinder to reach the outlet boundary. Following this decrease, the relative error subsequently increases for both cases, saturating around $15\%$ after approximately 10 advection times for $N_C=1$. For $N_C=2$, we have not been able to run the simulate for long enough to establish this saturation but envisage using a more efficient code implementation in the future to establish this result.

\begin{figure}[t!]
    \centering
    \includegraphics[width=0.9\linewidth]{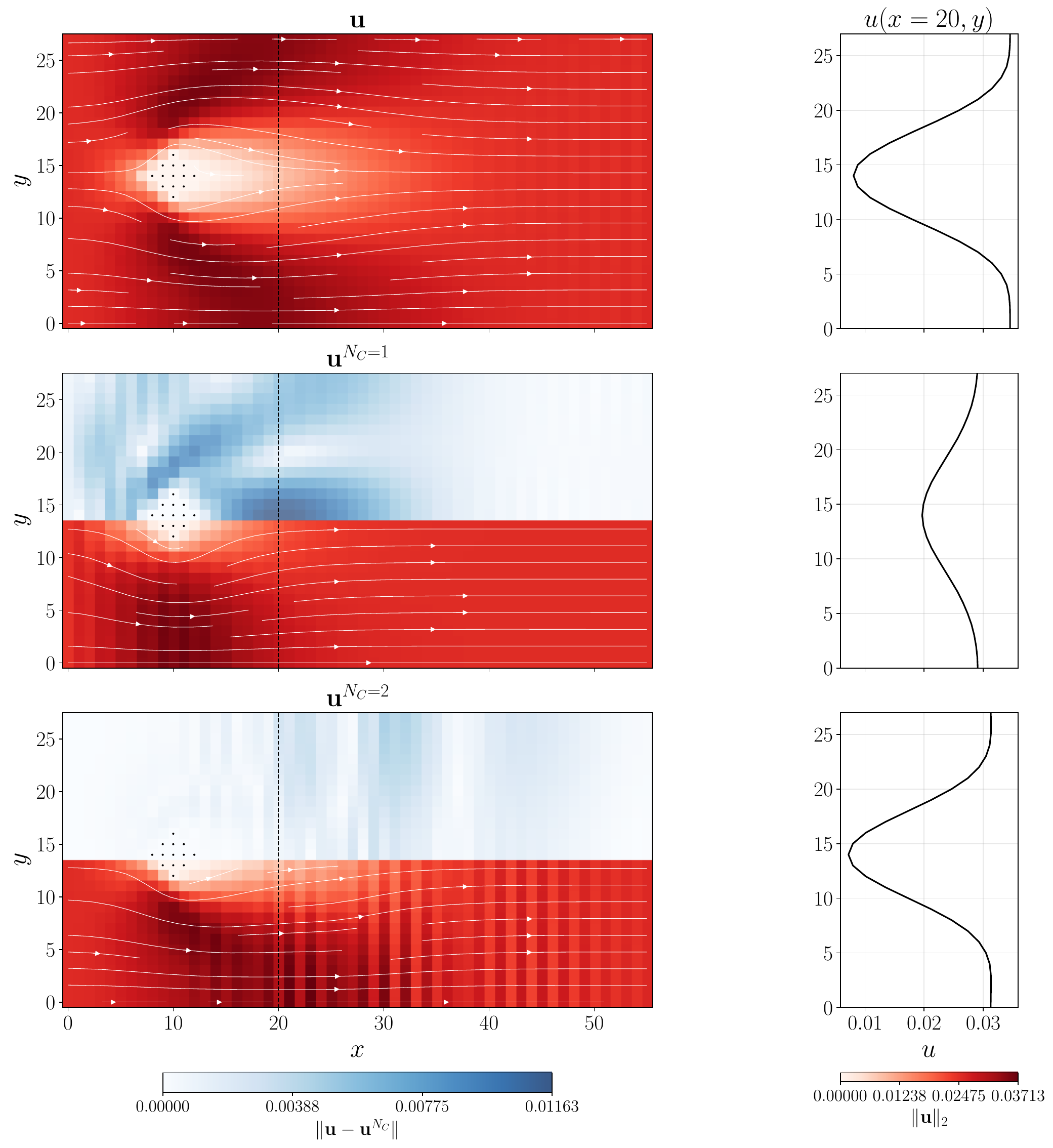}
    \caption{Spatial form of the classical LBM and Carleman LBM solutions (red) and their associated error (blue) for the flow around a cylinder after $5$ advection times for $\beta=0.75$ and $\re=6$. 
    Left: The classical LBM solution accompanied by its streamwise velocity profile (top), the linear Carleman LBM approximation (red, middle) and the quadratic Carleman LBM approximation (red, bottom). The Carleman LBM truncation error is shown in blue in the middle and bottom frames. Right: Horizontal velocity profile evaluated downstream of the cylinder at the location indicated by the dashed line.}
    \label{fig:cylinder_field_error}
\end{figure}

The top frame of Fig.~\ref{fig:cylinder_field_error} shows the classical solution obtained for this problem when $\re=6$ and $\beta=0.75$. Due to the computational constrains imposed by the Carleman LBM calculation, we use $N_x \times N_y = 58 \times 26$ lattice nodes, the cylinder is coarsely represented using $D=4$ (corresponding to $16$ lattice nodes), and we are restricted to steady state solutions. In this regime we observe higher velocities/kinetic energy as darker red above and below the cylinder. Behind the cylinder, the lower kinetic energy (lighter red) reflects the velocity deficit caused by the cylinder. The vertical profile of the horizontal velocity evaluated downstream of the cylinder, shown on the top right of Fig.~\ref{fig:cylinder_field_error}, also reflects this velocity deficit. 

The linear Carleman approximation (red) accompanied by its deviation from the classical LBM solution (blue) is shown in the middle of Fig.~\ref{fig:cylinder_field_error}. We observe that for the linear approximation the deviation is located primarily above and below the cylinder, where the increase in kinetic energy is misrepresented, as well as behind the cylinder, where the velocity deficit is misrepresented. The misrepresentation of the velocity deficit is emphasized in the deviation of $u(x=20,y)$, for $N_C=1$, from the classical simulation. Increasing the Carleman truncation order to $N_C=2$ (see bottom of Fig.~\ref{fig:cylinder_field_error}), we find that the misrepresentation of the classical LBM solution above/below and behind the cylinder is drastically reduced. This is also seen when comparing the profiles for $u(x=20,y)$ in the bottom and top right frames of Fig.~\ref{fig:cylinder_field_error}. By contrast, we observe that $N_C=2$  does not recover this velocity deficit without error, and in particular waves generated in the wake of the cylinder are found to be reflected from the outflow boundary in this case. Due to computational constraints it has not yet been possible to test $N_C=3$ in order to establish whether this is an error specific to $N_C=2$.


\subsection{Condition number}
\label{sec:condition}

As already mentioned in Sec.~\ref{sec:extensions}, the extensions to non-trivial geometries and driven flows introduced in this paper may bring additional complexity cost as compared to the original algorithm from Ref.~\cite{jennings2025end-to-end}, as these modifications may affect the condition number of the linear system from Eq.~\eqref{eq:AH_driving}. This, in principle, might negate the potential for a quantum advantage reported in Ref.~\cite{jennings2025end-to-end}. However, the numerical results that we will now present paint an optimistic picture, in which this extra cost is small and does not affect the asymptotic scaling of the algorithm's query complexity.

Due to the immense size of the linear system whose condition number we want to investigate (recall Table~\ref{table:dimensions}), we limit our numerical studies to $\beta=0.75$, $\re\in[5,20]$, $N_C=1,2$, and do not consider the flow past a cylinder case. Using a Lanczos-based algorithm, we first numerically compute the condition number $\kappa_{A_{\mathrm{per}}}$ of the linear system describing undriven and periodic LBE, i.e., ${A_{\mathrm{per}}}$ is given by Eq.~\eqref{eq:AH_driving} with Carleman driving $\D$ replaced by an identity matrix and Carleman streaming $\S$ constructed from the periodic streaming matrix given by Eq.~\eqref{eq:streaming_per}. Next, we compute $\kappa_{A_{\mathrm{vor}}}$ for ${A_{\mathrm{vor}}}$ describing the forced Taylor-Green vortex, i.e., again with periodic streaming, but this time with driving captured by Eq.~\eqref{eq:vor_driving}. Given that the Carleman driving matrix $\D$ is non-trivial only for $N_C\geq 2$ (as for $N_C=1$ driving is accounted for only by the Carleman driving vector $\F_0$), we only perform computations for $N_C=2$ (as for $N_C=1$ the results are identical to the undriven periodic case). Finally, we compute $\kappa_{A_{\mathrm{lid}}}$ for ${A_{\mathrm{lid}}}$ describing the lid-driven cavity flow, where now both the streaming and driving terms differ from the trivial case. As discussed above, for $N_C=1$, the driving has no effect and the difference between $\kappa_{A_{\mathrm{lid}}}$ and $\kappa_{A_{\mathrm{per}}}$ stems solely from a modified streaming matrix that now encodes bounce-back conditions at the walls enclosing the simulation region. Interestingly, we observe that in this case, up to numerical precision, $\kappa_{A_{\mathrm{lid}}}$ and $\kappa_{A_{\mathrm{per}}}$ are identical. This can potentially be attributed to the fact that the modification due to walls still preserves the unitary nature of the streaming matrix.

\begin{figure}[t!]
    \centering
    \includegraphics[width=0.49\linewidth]{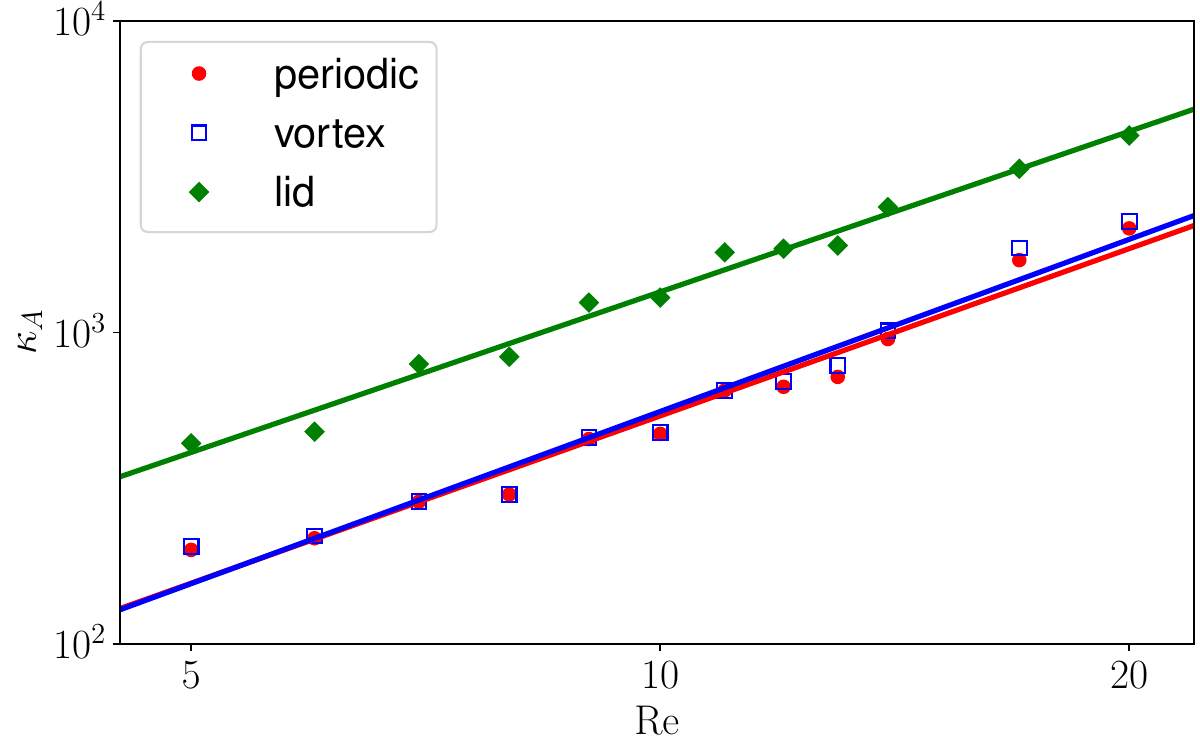}
    \caption{Condition number $\kappa_A$ as a function of the Reynolds number $\re$ for the periodic system, driven Taylor-Green vortex, and lid-driven cavity with parameters $\beta=0.75$ and $N_C=2$. Points correspond to numerical values and solid lines to best fits as described by Eq.~\eqref{eq:fits}.}
    \label{fig:condition}
\end{figure}

We present the results of our numerical studies for $N_C=2$ in Fig.~\ref{fig:condition}. To estimate the scaling of $\kappa_A$ with $\re$, we perform the best linear fits to:
\begin{equation}
    \label{eq:fits}
    \ln \kappa_A = \chi\ln\re+\ln c,
\end{equation}
yielding the scaling $O(\re^\chi)$. We extracted the following values of the exponents $\chi$ for the three cases investigated:
\begin{equation}
    \chi_{\mathrm{per}} = 1.784,\qquad\chi_{\mathrm{vor}} = 1.836,\qquad\chi_{\mathrm{lid}} = 1.712.
\end{equation}
Based on the above results, the following three observations can be made. First, given the relatively small number of data points and the fragility of doubly-logarithmic fits, the values of $\chi$ for the three considered cases are remarkably close to each other. This suggests that the algorithm's asymptotic query complexity for problems with non-trivial geometries and driving should not differ significantly from the simple case with no walls and periodic boundaries. Second, as illustrated by larger values of $\kappa_{\mathrm{lid}}$ as compared to $\kappa_{\mathrm{per}}$, there still may exist an extra resource cost, but one may expect it to be a constant overhead, not scaling with $\re$. Third, the classical complexity of solving a two-dimensional CFD problem scales as $O(\re^{2.25})$, which means that asymptotically our quantum LBM algorithm may outperform the best classical methods by a factor scaling roughly as $O(\re^{0.45})$. Even including the extra complexity cost $O(\re^{0.375})$ related to the extraction of the drag force from a quantum-encoded solution (as described in Ref.~\cite{jennings2025end-to-end}), should allow one to preserve a non-zero quantum advantage.


\section{Conclusion}
\label{sec:conclusions}

In this work, we have extended the recent quantum algorithm for the incompressible lattice Boltzmann method to include a class of flows and geometries that are much closer to those encountered in practical CFD, including walls, inlets, outlets, and body forcing. At the algorithmic level, we have shown how these ingredients can be incorporated into the Carleman streaming and driving operators, how they modify the linear system that underpins the quantum linear-solver--based algorithm, and what is the impact on the complexity of solving it. The resulting block-encodings for non-trivial geometries and driving lead to a controlled overhead, scaling as a multiplicative factor of order $4^{N_C}$ in the query complexity and a linear in the Carleman truncation order $N_C$ increase of ancillary qubits, which preserves the asymptotic scaling, and hence the window for a quantum advantage, identified in earlier work~\cite{jennings2025end-to-end}.

To assess the practical consequences of these extensions, we implemented a fully matrix-based classical Carleman LBM and benchmarked it against a standard incompressible LBM for three canonical two-dimensional flows: the driven Taylor--Green vortex, the lid-driven cavity flow, and flow past a cylinder. For the forced Taylor--Green vortex, we observed clear evidence of convergence of the Carleman approximation as the truncation order $N_C$ is increased, with $N_C = 3$ yielding a substantial reduction in error, and a relative error that saturates rather than grows in time. For the lid-driven cavity, the Carleman scheme with no-slip walls reproduces the main features of the recirculating flow and the associated shear layers. Moving from $N_C = 1$ to $N_C = 2$ significantly reduces the truncation error, even though the error grows with Reynolds number as the flow structure sharpens near the lid and corners. Finally, for flow past a cylinder at low Reynolds numbers, we demonstrated that moving from $N_C = 1$ to $N_C = 2$ again significantly reduces the truncation error for short times and that the same formalism can handle mixed inlet/outlet conditions and embedded obstacles. Note that a recent work~\cite{turro2025practical} has performed a related Carleman convergence study in a time-continuous lattice Boltzmann setting, i.e., simulating the discrete Boltzmann equation. This was done without the incompressibility assumption, with the Carleman truncation to second order, and for the Kolmogorov and lid-driven cavity flows. Our results are consistent with Ref.~\cite{turro2025practical} (in particular, in both works the relative error as a function of time plateaus), however they also extend and complement this work by including inlet/outlet boundaries and driving. 

From the perspective of quantum resources, our numerical study of the condition number of the Carleman linear system indicates that the inclusion of realistic boundaries and driving does not introduce a qualitatively new bottleneck, at least in the parameter ranges investigated. The dominant algorithmic costs remain those already identified for periodic, undriven flows, and the additional overhead from geometry and forcing enters mainly through the block-encoding prefactor and a modest increase in the number of ancillae. This supports the conclusion that, in regimes where the Carleman expansion converges, the end-to-end quantum LBM pipeline can be extended to more realistic CFD setups without closing the potential window for a quantum speedup. However, this window is currently far narrower than what was initially suggested in Ref.~\cite{li2025potential}, which claimed a potential for exponential quantum advantage. Therefore, it strongly motivates more work on pre-conditioning methods, problem-optimized block-encodings, and other algorithmic optimizations. Such advances could have a meaningful impact on the complexity scaling of these Carleman-based algorithms.

Our results also highlight important limitations and open questions. The present study is restricted to two-dimensional, low-Reynolds-number flows and relatively small Carleman orders $N_C \le 3$, dictated by the classical cost of simulating the embedded linear systems. The observed trends in truncation error with Reynolds number and geometry suggest that more complex flows, particularly three-dimensional turbulence, high-Reynolds-number flows, and strongly time-dependent configurations, will require both improved control of the Carleman convergence radius and more efficient representations of the embedded dynamics. Likewise, our analysis did not include observable extraction beyond the drag force already investigated in Ref.~\cite{jennings2025end-to-end}, which is an important line to push forward in future work.

Looking forward, several directions appear particularly promising. On the algorithmic side, it will be important to develop alternative linear embedding schemes adapted for particular CFD problems, preconditioning tailored to the LBM structure, and more economical block-encodings for boundary and forcing operators. On the application side, extending the classical numerical analysis to three dimensions, to higher-Reynolds-number benchmark flows, and eventually to industrially relevant geometries would provide sharper tests of whether practically meaningful quantum advantages can be realized in CFD. In particular, it would be of interest to extend classical simulations of the Carleman LBM implementation to Reynolds numbers at which the transition to time-dependent solutions and vortex shedding emerges, e.g., $\re \approxeq 83$ for the flow around a cylinder. This would test the ability of the Carleman method to accurately resolve such key physics. More broadly, our results illustrate that end-to-end quantum algorithm designs can be systematically pushed from idealized toy systems towards realistic engineering flows, and that they provide a concrete foundation for future studies at the interface of quantum computing and computational fluid dynamics.

\bigskip

\textbf{Authors contributions:} Authors are listed alphabetically within each affiliation. DJ contributed to the blocking-encoding theory. KK developed extensions to nontrivial flows and geometries, analyzed the corresponding complexity increases, and performed numerical simulations. ML supported the development of extensions to nontrivial flows and geometries. PM assisted in developing extensions to nontrivial flows and geometries, and performed numerical simulations. RA performed numerical studies of truncation errors and operator norms as input to complexity bounds. EM performed numerical comparison between LB and Carleman approximation. SR supported the definition of the end user requirements, and the application use cases. DJ, KK, and PM wrote the paper. 

\bigskip
	
\textbf{Acknowledgments:} We want to thank Thomas Astoul, Thomas Bendokat, Giovanni Giustini, and Alessandro De Rosis for helpful discussions concerning fluid dynamics and the lattice Boltzmann method. We are also grateful to Alice Barthe, Sam Morley-Short, and Trevor Vincent for their assistance with numerical simulations.


\section*{Appendix}

\subsection*{LBE discrete velocity models}
\vspace{0.5cm}

\begin{table}[ht]
\begin{center}
\begin{tabular}{|c|lll|}
    \hline
    \multirow{2}{*}{D1Q3} & $m=1$:&$\v{e}_m^\star=0$,& $w_m=2/3$, \\
    &$m\in\{2,3\}$:&$\v{e}_m^\star\in\{\pm 1\}$,&$w_m=1/6$.\\\hline
    \multirow{3}{*}{D2Q9} & $m=1$:& $\v{e}_m^\star=(0,0)$,&$w_m=4/9$, \\
    & $m\in\{2,\dots,5\}$: &$\v{e}_m^\star\in\{(\pm 1,0), (0,\pm 1)\}$,&$w_m=1/9$,\\
    & $m\in\{6,\dots,9\}$:& $\v{e}_m^\star\in\{(\pm 1, \pm 1)\}$,& $w_m=1/36$.\\ \hline
    \multirow{4}{*}{D3Q27} & $m=1$:& $\v{e}_m^\star=(0,0,0)$,& $w_m=8/27$, \\
    & $m\in\{2,\dots,7\}$: & $\v{e}_m^\star\in\{(\pm 1,0,0), (0,\pm 1,0),(0,0,\pm 1)\}$,& $w_m=2/27$,\\
    & $m\in\{8,\dots,19\}$:& $\v{e}_m^\star\in\{(\pm 1, \pm 1,0),(\pm 1, 0,\pm 1),(0,\pm 1, \pm 1)\}$,& $w_m=1/54$,\\
    &$m\in\{20,\dots,27\}$:& $\v{e}_m^\star\in\{(\pm 1, \pm 1,\pm 1)\}$,& $w_m=1/216$.\\\hline
\end{tabular}
\caption{Discrete velocities in lattice units for standard LBE models~\cite{kruger2016lattice}.}
\label{table:D2Q9}
\end{center}
\end{table}

\subsection*{Streaming and collision matrices}

The streaming matrix $S$ can be explicitly written as:
\begin{equation}
    \label{eq:streaming}
    S = \sum_{\v{r}^\star}\sum_{m=1}^Q \ketbra{\v{r}^\star+\v{e}_m^\star}{\v{r}^\star} \ot \ketbra{m}{m},
\end{equation}
whereas the linear and quadratic collision matrices, $F_1$ and $F_2$, have the following form:
\begin{aligns}
    \label{eq:F1}
    &F_1 =  \sum_{\v{r}^\star} \ketbra{\v{r}^\star}{\v{r}^\star} \ot \sum_{m,m_1=1}^Q  \frac{1}{\tau^\star+\frac{1}{2}} (w_m+3 w_{m} E_{m,m_1} - \delta_{m,m_1})\ketbra{m}{m_1},\\[12pt]
    \label{eq:F2}
    &F_2 = \sum_{\v{r}^\star} \ketbra{\v{r}^\star}{\v{r}^\star, \v{r}^\star} \ot \sum_{m,m_1,m_2=1}^Q \frac{w_m}{\tau^\star+\frac{1}{2}}  \left(\frac{9}{2}E_{m,m_1}E_{m,m_2}-\frac{3}{2}E_{m_1,m_2}\right)\ketbra{m}{m_1,m_2},
\end{aligns}
where $E_{m,n}:= \v{e}^\star_m \cdot \v{e}^\star_n$ and $\delta$ is the Kronecker delta.

\subsection*{Carleman streaming and collision matrices}

The Carleman streaming matrix $\S$ for truncation order $N_C$ is given by the following block-diagonal matrix:
\begin{equation}
    \label{eq:embeded_str_bound}
    \mathcal{S}^\infty = 
    \begin{pmatrix}
        S & 0 & \dots & 0\\
        0     & S^{\otimes 2} & \dots & 0\\
        \vdots&\vdots & \ddots& \vdots \\
        0     & 0   &\dots & S^{\otimes N_C}\
    \end{pmatrix}.
\end{equation}
The Carleman collision matrix $\C$ for truncation order $N_C$ has the following form:
\begin{equation}
\C = 
\begin{pmatrix}
    C^1_1 & C^1_2 & 0 & 0     &\dots&0\\
    0     & C^2_2 & C^2_3 & C^2_4 &\dots&0\\
    0     & 0     & C^3_3 & C^3_4 &\dots&0\\
    \vdots&\vdots & \vdots& \vdots& \ddots &\vdots\\
    0 & 0 & 0 & 0 & \dots & C_{N_C}^{N_C} 
\end{pmatrix}, 
\end{equation}
where in each row $k$ there are $\min\{k+1,N_C-k+1\}$ non-zero blocks given by $d^k\times d^l$ matrices:
\begin{equation}
\label{eq:Ckl}
    C_l^k:=\left[\left( I+F_1\right)^{\otimes 2k-l} \otimes F_2^{\otimes l-k}+\mathrm{perms.}\right],
\end{equation}
with $+ \ \mathrm{perms.}$ denoting a sum over distinct permutations over $k$ subsystems, with $2k-l$ subsystems of type 1, and $l-k$ subsystems of type 2.

\subsection*{Numerical validation}

\begin{figure}[h]
    \centering
    \includegraphics[width=0.49\linewidth]{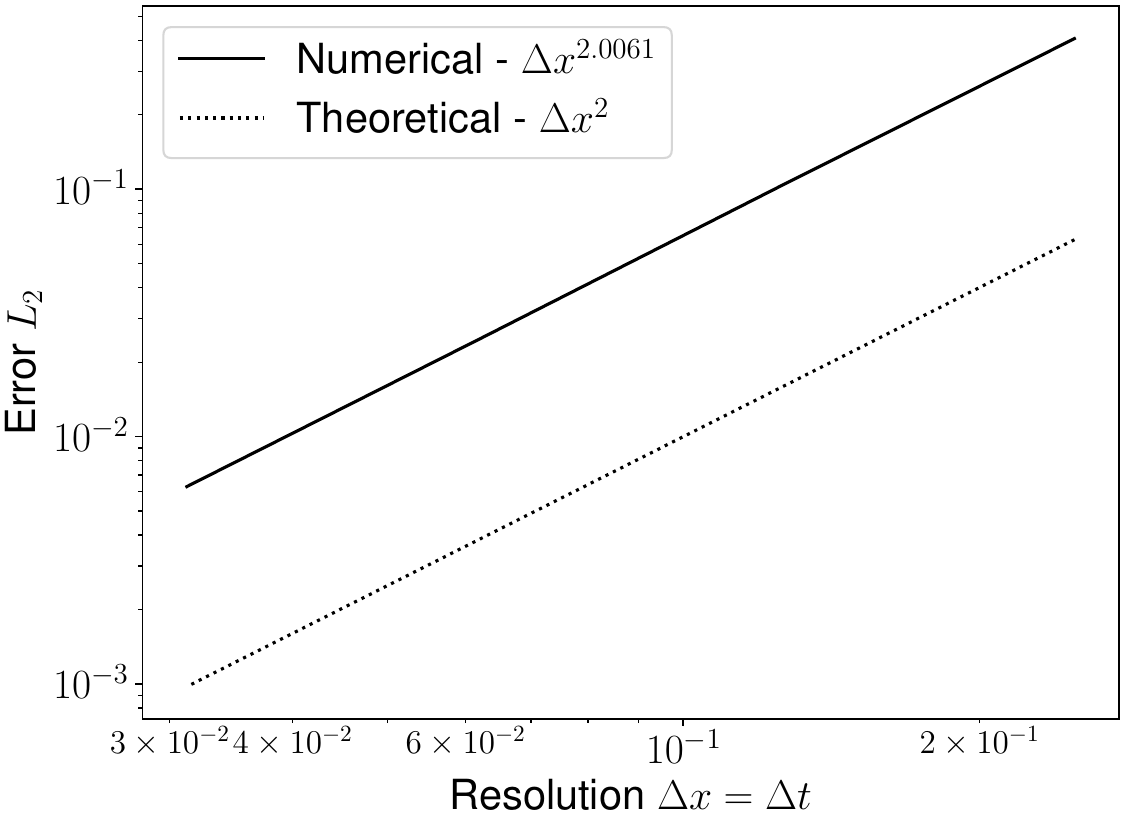}
    \includegraphics[width=0.49\linewidth]{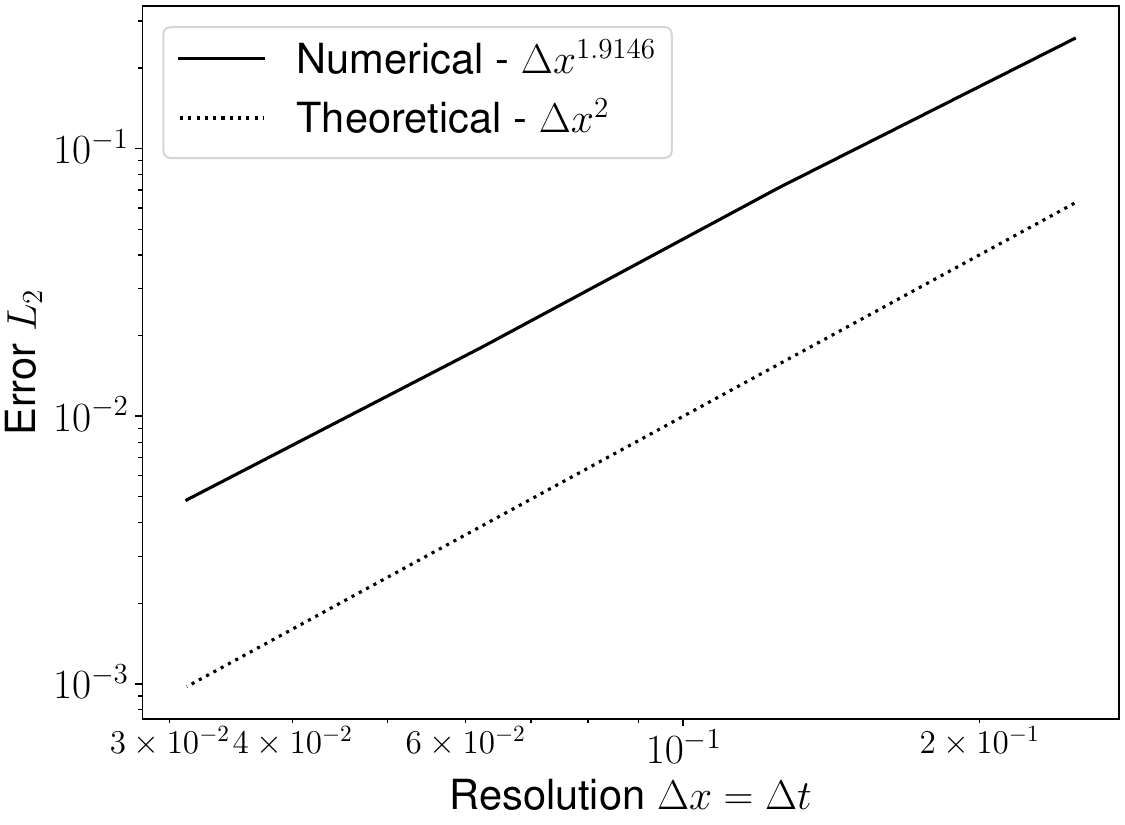}
    \caption{Validation that our code for $\re=10, u^\star_0 = 0.5$ correctly reproduces (left) the time evolution of the analytical solution for the decaying Taylor-Green vortex and, (right) the analytical solution for the forced Taylor-Green vortex with second-order accuracy in both space and time. Although the forcing used is first-order accurate, the dynamics are steady and we almost recover second order temporal accuracy.}
    \label{fig:TG_Convergence_Validation}
\end{figure}

\bibliography{references}

\end{document}